\documentclass[a4paper,11pt]{scrartcl}
\usepackage{latexsym}
\usepackage{epsfig}
\usepackage{amsmath}
\usepackage{amsfonts}
\usepackage{amssymb}
\usepackage{subfigure}
\usepackage[usenames]{color}
\usepackage{pgf}
\input{epsf}
\usepackage{graphicx,amssymb,amsmath,amsfonts,bm}
\usepackage[latin1]{inputenc}
\usepackage[T1]{fontenc}
\usepackage{cite}
\usepackage{epsfig}
\usepackage{graphicx}
\usepackage{epsfig}
\usepackage{dcolumn}
\usepackage{amsfonts}
\usepackage{float}
\usepackage{amsthm}
\usepackage{epstopdf}

\newtheorem{thm}{Theorem}[section]

\newtheorem{lem}[thm]{Lemma}

\def\ii{{\rm i}} % roman i for use as sqrt(-1)
\def\ee{{\rm e}} % roman e for use as exp
\def\dd{{\rm d}} % roman d for use as derivative
\def\ub{{\boldsymbol{u}}}

\def\Ub{{\boldsymbol{U}}}
\def\xb{{\boldsymbol{x}}}
\def\Tb{{\boldsymbol{T}}}
\newcommand{\taub}{\bm{\tau}}
\newcommand{\gammadot}{\dot{\gamma}}
\newcommand{\gammabdot}{\bm{\dot{\gamma}}}
\newcommand{\elast}{\epsilon\ell}

\newcommand{\Hhat}{\overline{\mathcal{H}}}
\def\grad{{\nabla}}
\def\gradb{\bm{\nabla}}
\newcommand{\ddy}[2]{\frac{\partial#1}{\partial#2}} 
\newcommand{\dy}[1]{\partial#1} 
\newcommand{\lapl}{\Delta}

\newcommand{\trademark}[1]{#1\textsuperscript{\textregistered}}

\title{Linear Stability Analysis for Plane-Poiseuille Flow of an Elastoviscoplastic fluid with internal microstructure}
\author{Miguel Moyers-Gonz\'alez$^a$\thanks{Corresponding author: miguel.moyersgonzalez@canterbury.ac.nz}, Teodor I. Burghelea$^b$ and Julian Mak$^c$\\
\small $^a$Department of Mathematics and Statistics \\
\small University of Canterbury, Private Bag 4800, Christchurch, 8041, NZ \\
\small $^b$Institute of Polymer Materials\\
\small Friedrich-Alexander-University Erlangen-Nurnberg, D-91058 Erlangen, Germany\\
\small $^c$School of Mathematics\\
\small University of Leeds, Leeds, LS2 9JT, UK
}

\begin{document}

\maketitle

\hrule
\begin{abstract}
We study the linear stability of Plane Poiseuille flow of an elastoviscoplastic fluid using a revised version of the model proposed by Putz and Burghelea (Rheol. Acta (2009)48:673-689). The evolution of the microstructure upon a gradual increase of the external forcing is governed by a structural variable (the concentration of solid material elements) which decays smoothly from unity to zero as the stresses are gradually increased beyond the yield point. Stability results are in close conformity with the ones of a pseudo-plastic fluid. Destabilizing effects are related to the presence of an intermediate transition zone where elastic solid elements coexist with fluid elements. This region brings an elastic contribution which does modify the stability of the flow. 

\end{abstract}

\hrule

%%%%%%%%%%%%%%%%%%%%%%%%%%%%%%%%%%%%%%%%%%%%%%%%%%%%%%%%%%%%%%%%%%%%%%%%
%%%%%%%%%%%                                                  Introduction                                                                                                      %%%%%%%%%%%%%%
%%%%%%%%%%%%%%%%%%%%%%%%%%%%%%%%%%%%%%%%%%%%%%%%%%%%%%%%%%%%%%%%%%%%%%%%
\section{Introduction}

During the past several decades, yield stress fluids found an increasing number of practical applications for several major industries (which include cosmetics, foods, oil field etc.) and they are encountered in the daily life in various forms such as hair gels, food pastes, cement, mud and more. Such materials can be loosely and simplistically defined as materials that do not flow unless a minimal stress (referred as 'yield stress') is applied onto them. Understanding and controlling the hydrodynamic stability during flows of yield stress fluids is important for practical applications which involve flows of such materials through conduits.
From a fundamental standpoint, yield stress materials continue triggering intensive debates and posing difficult challenges, of both theoretical and experimental nature. Undoubtedly, the best known and most cited debate is related to the very definition of such materials and the existence of a 'true yield stress', \cite{1985Barnes_tysam,1999Barnes_tysar}. It was argued in Refs. \cite{1985Barnes_tysam,1999Barnes_tysar} that the yield stress emerges as an artefact related to the inability of the rheometric equipment to properly identify a viscous flow regime in a range of negligibly small rates of deformation. Recent studies claim to have solved the "yield stress debate" by arguing that prior to yielding the viscosity of the material is infinite which demonstrates the existence of a true solid state,\cite{bonn_europhys, bonn_science}.
\subsection{Yielding of a \trademark{Carbopol} gel and its relevance to hydrodynamic studies}
 The existence of a "true" yield stress and the nature of the transition from a solid to fluid behaviour may look at a first glance of little or no relevance to the hydrodynamic stability of the flow of a yield stress material. Indeed, whereas the yielding transition occurs at low values of the Reynolds number (typically $Re<1$) a loss of the hydrodynamic stability is typically observed at significantly larger $Re$ (typically $Re>1000$). On the other hand, the base flows usually considered in the linear analysis of the hydrodynamic stability of channel flows of yield stress fluids are characterized by a significant stratification of the velocity gradients: large values near the channel boundaries which are consistent with a yielded flow region and vanishing values near the center-line, which are consistent with a plug region. A recent experimental investigation of the laminar-turbulent transition in the pipe flow of a yield stress fluids demonstrates that the transition to turbulence occurs when the Reynolds stresses balance the yield stress of the fluid, that is when the solid plug is broken, \cite{bulent_crap}. These findings corroborate well with the idea that, in fact, the nature of the solid-fluid transition and the yielding scenario may play an important role in the stability problem. A rather new and certainly unexpected insight into the yielding transition has been brought by an experimental study on sedimentation of spherical particles in a physical gel (\trademark{Carbopol} $940$), \cite{sedimentation_andreas}. Quite unexpectedly, the fore-aft symmetry of the flow patterns around the falling object was broken (in spite of the smallness of the Reynolds number during these experiments) and even more surprisingly a negative wake has been observed (see Fig. $8$ in Ref. \cite{sedimentation_andreas}). A similar non symmetric flow pattern has been observed experimentally during flows of \trademark{Carbopol} gel past a cylinder, \cite{Tokpavi200935}.
A first message of the sedimentation study presented in \cite{sedimentation_andreas} was that the solid-fluid transition in the \trademark{Carbopol} gel might not be reversible upon increasing/decreasing stresses, unlike it has been traditionally assessed. A second important message was that the elastic effects responsible for the emergence of the experimentally observed negative wake might be consistent with an intermediate transition regime where solid material elements (unyielded) coexist with fluid (yielded) ones. A more systematic investigation of these features of the solid-fluid transition in a \trademark{Carbopol} gel has been presented 
in Ref. \cite{Burghelea09}. Indeed, as suggested by the sedimentation experiments presented in \cite{sedimentation_andreas}, a hysteresis has been found in the solid-fluid transition of a \trademark{Carbopol} gel upon increasing/decreasing stresses (see Fig. 3 in Ref. \cite{Burghelea09}). A clear signature of the elasticity involved in the transition region in the form of a recoil effect (negative shear rates) on the decreasing stress branch of the flow curve has been observed as well.    
  The hysteresis effect observed in the increasing/decreasing stress flow curves and reported in \cite{Burghelea09} has also been observed experimentally by Divoux et al., \cite{manneville}, and attributed to critical slowing down of the system dynamics close to yielding (and not to any memory effect).
\subsection{Linear stability analysis of a yield stress fluid: a brief survey}

The first study on linear stability analysis of a Bingham viscoplatic fluid is due to Frigaard et al. \cite{Frig94}. These authors concentrated on the asymptotic stability of two-dimensional traveling wave perturbations and found that the critical Reynolds number increases linearly with Bingham number. They also showed that the plug region remains unaffected by the disturbance field. In this work the authors imposed even symmetry at the yield surface and the results are incomplete. Some years later Nouar et. al. studied again this problem using the exact boundary conditions at the yield surface and they found that the plane-Bingham-Poiseuille flow is linearly stable. This being a consequence of the vanishing perturbation at the yield surface. They also conjectured that this result could be extended to Poiseuille flow of a viscoplastic fluid in a circular or annular pipe and to other rheological models such as the Herschel-Bulkley and Casson models.

As mentioned before, recent studies have found that the solid-fluid transition in some yield stress materials, such as \trademark{Carbopol}, is not reversible under increasing/decreasing stresses, thus the objective of this work is the linear stability analysis of an elastoviscoplastic fluid.
In this work we consider a modified version of the structural model developed by Putz and Burghelea in \cite{Burghelea09}. This models consists of a non-linear Maxwell-type viscoelastic constitutive equation and a kinematic equation that governs the behaviour of the microstructure. The relaxation time of the fluid depends on the structural variable which is equal to one if the fluid is unyielded and zero if the fluid is fully yielded. We should note that the transition from viscoelastic solid to fluid is continuous and smooth. For the non-linear viscosity we consider a regularized viscoplastic model, its behaviour is the one of a pseudo-plastic fluid, that is, if the second invariant of the stress tensor is below the yield stress the fluid presents a very high viscosity.

Several authors have studied the linear stability problem of a flow with variations in viscosity across the channel \cite{Chikkadi05,Nouar07,Pinarbasi01}. Govindarajan et. al. \cite{Govindarajan01} showed that by having a viscosity which is decreasing function of space it is sufficient to considerably delay the onset of two-dimensional instabilities. This result was latter confirmed by Nouar and Bottaro for the case of shear-thinning fluids \cite{Nouar07}. Linear stability analysis for a regularized Bingham model were carried out by Frigaard and Nouar in \cite{Frig05}. They showed that the spectrum for the regularized problem had some physical spurious eigenvalues, not present in the Bingham problem. The stability of these eigenvalues depended on the size of the regularization parameter and unstable modes can be found even for moderate Reynolds numbers if this parameter is small enough.

For the case of Maxwell liquids, Gorodstov and Leonov \cite{Gorod67} showed the existence of a stable continuous spectrum and two stable eigenvalues for each value of the wave number in the streamwise direction, $\alpha$. In the same work they predicted instabilities at finite Reynolds numbers. Later, this result was proven wrong by several authors, e.g.~ \cite{Denn77,Lee86,Renardy86}. Through careful numerical simulation of the generalised eigenvalue problem no unstable modes were found, even for high Reynolds numbers. Denn and coworkers \cite{Denn77,Denn72,Denn73} also studied the stability of plane Poiseuille flow and found that at high Reynolds numbers UCM fluids are significantly less stable than Newtonian fluids. Sureshkumar \& Beris \cite{Suresh95} confirmed this and also showed that destabilisation is reduced for the Oldroyd-B and Chilcott-Rallison models. Renardy \cite{Rena92} has proven the linear stability of Couette flow of a UCM fluid, but there is no proof of linear stability of Couette flows for more general fluids of this type. However, it is generally believed that plane Couette flow is linearly stable. Indeed much of the literature is focused at the study of interesting features of the eigenspectra, rather than marginal stability, e.g.~\cite{Wils99,Kupf05}.

%%%%%%%%%%%%%%%%%%%%%%%%%%%%%%%%%%%%%%%%%%%%%%%%%%%%%%%%%%%%%%%%%%%%%%%%
%%%%%%%%%%%%%%%                 Model, and experiments 											%%%%%%%%%%%%%
%%%%%%%%%%%%%%%%%%%%%%%%%%%%%%%%%%%%%%%%%%%%%%%%%%%%%%%%%%%%%%%%%%%%%%%%

%---------------------------------
\section{Elastoviscoplastic model with internal microstructure}\label{sec_pmm}
The elastovisplastic fluid is described by the following set of equations with unknowns $(p,\ub,\taub,\Phi)$:
\begin{eqnarray}
\rho\left(\frac{\partial \bm{u}}{\partial t}+(\bm{u}\cdot{\bm \nabla})\bm{u}\right)&=&-\nabla p+\mu_s {\bm \nabla} \cdot \bm{\dot{\gamma}}+ {\bm \nabla} \cdot\bm{\tau} \label{eqnmotiondim}\\
{\bm \nabla} \cdot\bm{u}&=& 0 \label{massconsdim},
\end{eqnarray}
where $\mu_s$ denotes the solvent viscosity. The rate of strain tensor is $\bm{\dot{\gamma}} = \bm{\dot{\gamma}}({\bm u})$.
The constitutive equation for the elastic stress tensor $\bm{\tau}$ is:
\begin{equation}
\bm{\tau}+\frac{\mu\left(\dot{\gamma}(\bm{u})\right)}{G} \Phi \stackrel{\triangledown}{\bm{\tau}} = \mu\left(\dot{\gamma}(\bm{u})\right) \bm{\dot{\gamma}}, \label{constdim}
\end{equation}
where
\[
\stackrel{\triangledown}{\cdot}=\frac{D \cdot}{D t}-\bm \nabla\bm{u}\cdot-\cdot \bm \nabla\bm{u}^T
\]
is the Upper-Convected-Derivative and $D\cdot /Dt$ is the usual material derivative.
Finally, the concentration of the solid state, $\Phi$, satisfies the following kinematic equation:
\begin{equation}
\frac{\partial \Phi}{\partial t}+(\bm{u}\cdot{\bm \nabla})\Phi=R_d(\Phi,\tau(\bm{u}))+R_r(\Phi,\tau(\bm{u})),\label{phieqndim}
\end{equation}
with $R_d$ being the rate of destruction of solid units and $R_r$ is the rate of fluid recombination of fluid elements into a gelled structure.

Note that when $\Phi\equiv 1$ the model describes a Maxwell type viscoelastic fluid with a nonlinear viscosity $\mu(\gammadot)$ and if $\Phi\equiv 0$ the model represents a Generalized Newtonian fluid.
 
In both (\ref{constdim}) and (\ref{phieqndim}), $\dot{\gamma}$ and $\tau$ are the second invariant of the rate of strain and elastic stress tensors respectively. These are defined as follow:
\[
\dot{\gamma}(\bm{u})=\left(\frac{1}{2}\sum_{i,j=1}^3[\dot{\gamma}_{ij}(\bm{u})]^2\right)^{1/2},~~~~~~\tau(\bm{u})=\left(\frac{1}{2}\sum_{i,j=1}^3[\tau_{ij}(\bm{u})]^2\right)^{1/2}.
\]
%------------------------------------------------------------------------------
%---------------------------
%-----------------------------------------------------------------------------
\subsection{Choice of functions for nonlinear viscosity $\mu\left(\dot{\gamma}\right)$, destruction rate $R_d(\Phi,\tau(\bm{u}))$ and recombination rate $R_r(\Phi,\tau(\bm{u}))$.}

In this section we present our choices for the nonlinear viscosity $\mu(\gammadot)$ and the destruction and recombination rate functions. The choices are clearly non unique. Constitutive models for fluids with a yield stress trace back to the early 1900's . The most widely used are the ones due to Bingham \cite{Bingham}, Herschel-Bulkley \cite{H-B} and Casson \cite{Casson}. Description of the rates of destruction and recombination is rather empirical and for this we follow closely \cite{Burghelea09}.

%-------------------------------------

\subsubsection{Non-linear viscosity $\mu(\dot{\gamma})$}

The fluids we are interested in present a yield stress. It is well known that an ``ideal'' viscoplastic fluid behaves as a Newtonian (or Generalized Newtonian) fluid if $\tau$ is above the yield stress, if the contrary happens then the material behaves as a plastic solid and one says that the fluid is unyielded. In \cite{Burghelea09} it has been shown that the transition from an elastic solid behavior (unyielded) to a fluid behavior (yielded) is not direct but mediated by a viscoelastic like regime where solid and fluid behavior coexist. This effect, though previously unnoticed, is particularly pronounced in the case of "fast" yielding, i.e when the externally applied stress changes unsteadily and becomes vanishingly small in the limit of steady yielding, in other words, when the stresses vary infinitesimally slow. Although it has been argued that if one waits ``long enough'' the viscosity in the unyielded region actually tends to infinity, see \cite{bonn_europhys}, in an experimental (to investigate hydrodynamic stability for example) or industrial setting it is unlikely one will reach those time frames. Therefore for the rest of this study we will work with a regularized version of a viscoplastic constitutive model.  Because our main interest is on linear stability analysis and this involves high Reynolds numbers it is better if we have a model with non-zero infinite shear viscosity, for this reason we choose the shear-thinning regularized Casson's model for the viscosity:
\begin{equation}
\mu(\dot{\gamma})=\left(\mu_{\infty}^{1/m}+\left(\frac{\tau_y}{\sqrt{\dot{\gamma}^2+\epsilon^2}}\right)^{1/m}\right)^m,\label{Cassonvisc}
\end{equation}
where $m\geq 1$ is the power law index and $\tau_y$ is the yield stress and $\epsilon\ll 1$ is the regularization parameter . We recover Casson's model when $m=2$, Bingham model when $m=1$ and the model reduces to a Newtonian viscosity when $m=1$ and $\tau_y=0$. In Figure \ref{fig:viscfandg}a show an example of (\ref{Cassonvisc}), note that if $\tau<\tau_y$ the viscosity becomes very large but never infinity as long as $\epsilon$ is fixed and different from zero, thus  we will be working with a pseudo-plastic viscosity (shear-thinnig). 

%--------------------------------------------
\subsubsection{$R_d(\Phi,\tau(\bm{u}))$ and $R_r(\Phi,\tau(\bm{u}))$}

\begin{figure}
\begin{center}
(a)\scalebox{0.4}{\includegraphics{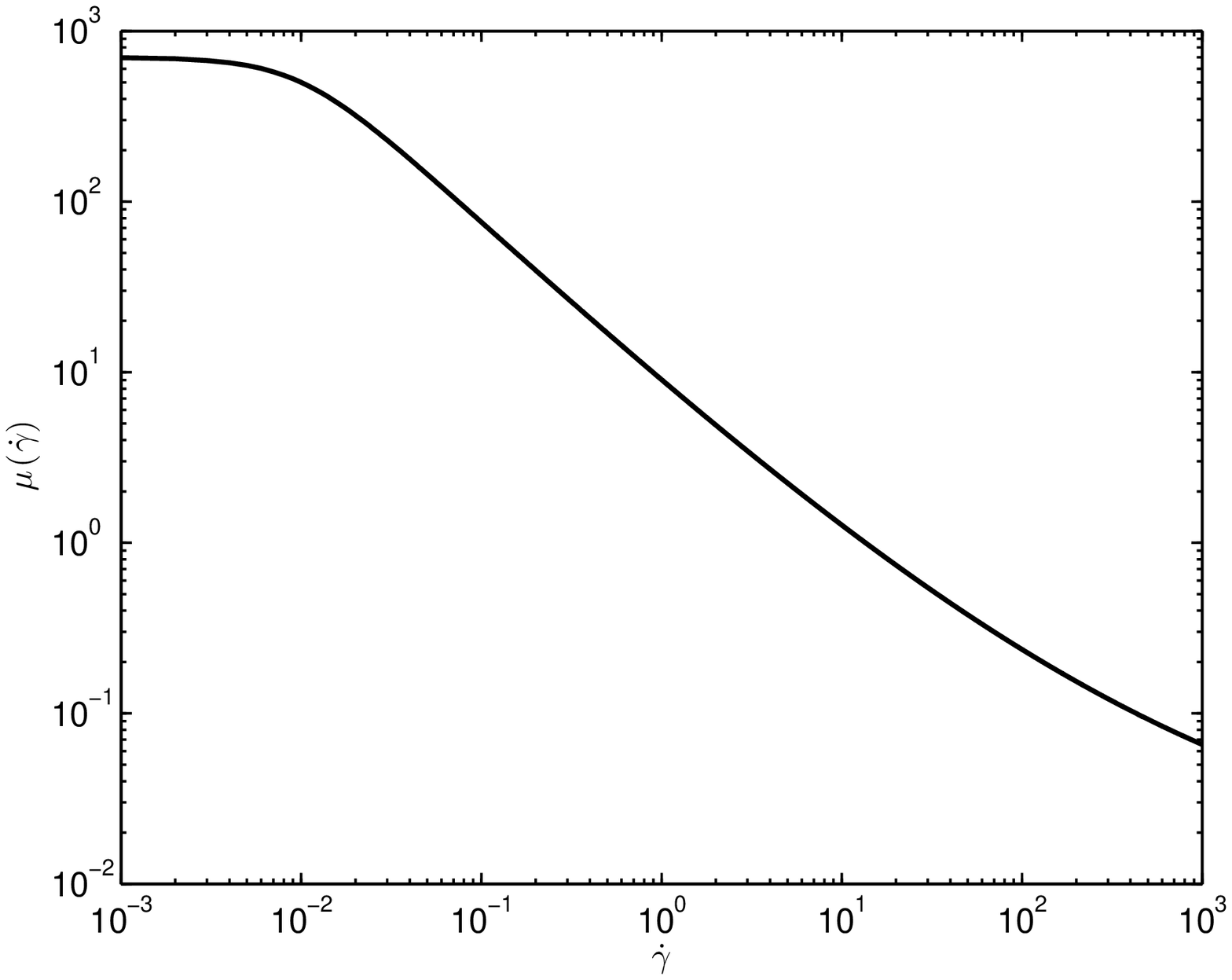}}
(b)\scalebox{0.4}{\includegraphics{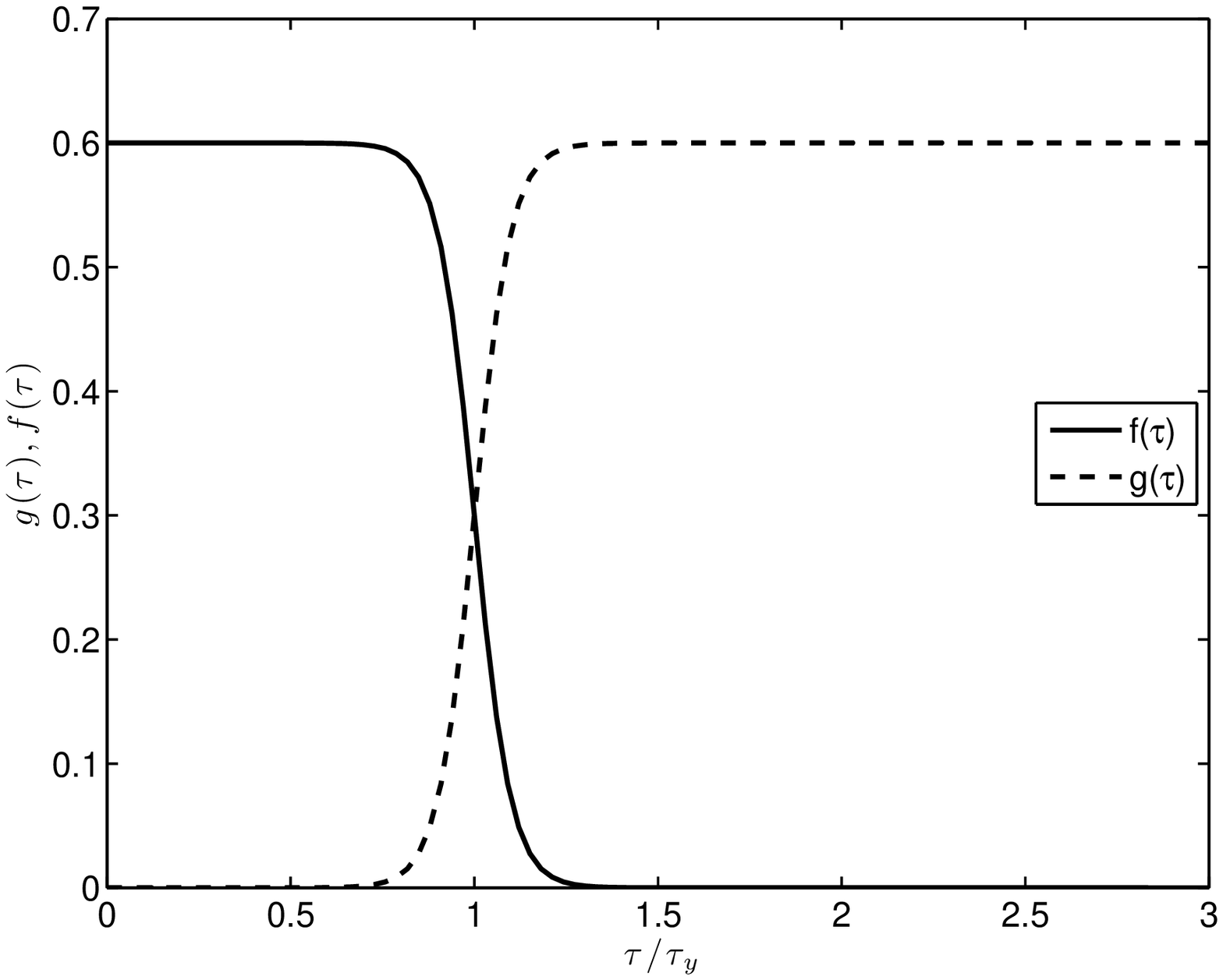}}
\end{center}
\caption{(a) Viscosity function $\mu(\dot{\gamma})$ with $m=3$, $\mu_{\infty}=0.0102$, $\tau_y=6.5$ and $\epsilon=0.01$. (b) $f(\tau)$ and $g(\tau)$ with $K_d=K_r=0.3$ and $w=0.1$.}
\label{fig:viscfandg}
\end{figure}

As discussed by Putz and Burghelea in \cite{Burghelea09} $R_d$ and $R_r$ have to satisfy the following assumptions:
\begin{itemize}
\item[1.] $R_d(\Phi,\tau(\bm{u}))$ is proportional to the relative speed of neighboring solid units and the existing amount of solid, thus
\begin{equation}
R_d(\Phi,\tau(\bm{u}))=-g(\tau(\bm{u}))\Phi,
\end{equation}
and
\begin{equation}
g(\tau(\bm{u}))=K_d\left(1+\tanh\left(\frac{\tau(\bm{u})-\tau_y}{w}\right)\right),\label{geqn}
\end{equation}
where $g(\tau)$ is an increasing function of the second invariant of the stress tensor, $K_d$ is the rate of destruction with units $s^{-1}$ and $w$ determines the ``width'' of the solid-fluid region and has units Pa.
\item[2.] $R_r(\Phi,\tau(\bm{u}))$ decreases with the relative speed of neighboring elements, begin practically zero in a fast enough flow. Therefore,
\begin{equation}
R_r(\Phi,\tau(\bm{u}))=f(\tau(\bm{u}))(1-\Phi),
\end{equation}
and
\begin{equation}
f(\tau(\bm{u}))=K_r\left(1-\tanh\left(\frac{\tau(\bm{u})-\tau_y}{w}\right)\right),\label{feqn}
\end{equation}
\end{itemize}
where $f(\tau)$ is a decreasing function of the second invariant of the stress tensor. In Figure \ref{fig:viscfandg}b we show an example of (\ref{geqn}) and (\ref{feqn}). Note that when $\tau$ is far below the yield stress, the material is fully ``unyielded'' which means it behaves as a viscoelastic solid. This because when $\tau<\tau_y$, $\mu(\gammadot)$ is of the order $O(1/\epsilon)$ and $\Phi\equiv 1$, thus by a simple dominant balance analysis equation (\ref{constdim}) reduces to the one for a Kelvin-Voight viscoelastic solid. If the opposite happens ($\tau$ far above the yield stress) $\Phi\equiv 0$ and the fluid behaves as a shear-thinning fluid. It is important to note  that in a neighborhood around $\tau_y$ the material is neither a viscoelastic solid nor a viscous fluid, both solid and fluid structures coexist and this region will play a crucial role int the stability of the flow. The existence of this region is the main difference with respect to other proposed models for this type of elastoviscoplastic fluids. The main difference of the approach used here and previous work, see \cite{Moller06}, is that we do not use the steady states of \eqref{phieqndim} but solve it numerically and thus describe a non-steady yielding behavior which we believe might be of some practical relevance 

%-----------------------------------------------------------------

\subsection{Validation of the model against experimental results}\label{sec_expcomparison}

To validate the model, we use the rheological measurements presented in \cite{Burghelea09} conducted for various \trademark{Carbopol} solutions with concentrations (by weight) ranging in between $0.05 \%$ and $0.2 \%$.
The measurements were conducted on a Bohlin (now Malvern) $C-VOR$ rotational rheometer equipped with a
digitally controlled temperature (within $0.1 ^\circ C $ ) bath
($RTE-111$ from Neslab). The rheometer was operated in a controlled stress mode. To insure a good reproducibility of the measurements,
prior to each experiment the sample has been pre-sheared at
a constant stress (usually the largest stress applied during the
test, so that yielding occurs in the entire volume of the material) for $300~s$. After the pre-shear step, the samples were allowed to relax for $300~s$ prior to the rheological measurements.
 A first major concern during these experiments was the occurrence of the wall slip phenomenon at the
contact with the measuring geometry, which is a well known and documented effect for \trademark{Carbopol} gels, \cite{piau}. To prevent this, a serrated parallel plates geometry with an
approximate roughness of $0.8~mm$ has been used. A second concern was related to
the possible artefacts introduced by fluid evaporation during long
experimental runs. To prevent this, a solvent trap has
been placed around the free fluid meniscus. After each
experimental run we have carefully checked (by visual inspection)
that no significant changes in the shape of the meniscus occurred,
and thus concluded that evaporation effects were either minimal or
absent. The radius of the plates was $R=2~cm$ and the distance
between them was $d=1~mm$ (measured between the plates
protuberances). Two types of rheological tests have been
performed: controlled stress linear ramps and controlled stress
oscillatory sweeps at a fixed frequency. For each sample we have
conducted controlled stress experiments for $19$ different values
of the total ramp time. Each constant stress rheological
experiment started with a increasing stress ramp and ended with
a decreasing stress ramp within the same range of stresses and the
same stress step. The data averaging time per stress value, $t_0$ (referred in \cite{Burghelea09} as the \textit{characteristic time of
forcing}), has been varied between $0.2~s$ and $2~s$. For each
up-down stress ramps $1000$ stress values have been explored
ranging in between $0.1~Pa$ and $20~Pa$. 
We emphasize here that a true steady state of deformation can only be inferred in the asymptotic 
limit $t_0 \rightarrow \infty$, unlike in previous studies concerning \trademark{Carbopol} gels where a steady state was a priori set by deliberately choosing large values of $t_0$ (an accurate description of such procedure is presented, for example,  in Ref. \cite{coussotjnfm2008}). 
Alternatively, the time dependent response of the samples has been tested via stress
controlled oscillatory experiments at several frequencies and in
the same range of stresses.

We now turn our attention to the comparison of the predictions of the model described in Sec. \ref{sec_pmm} with the experimental results presented in Ref. \cite{Burghelea09}.

Bearing in mind that the experiments provide a scalar data set, the equations (\ref{eqnmotiondim})-(\ref{phieqndim}) reduce to:

\begin{eqnarray}
&&\frac{d\Phi}{dt}=-g(\tau)\Phi+f(\tau)(1-\Phi)\label{eqn:phiexperiment}\\
&&\frac{\mu(\dot{\gamma})}{G}\Phi\frac{d\tau}{dt}+\tau=\mu(\dot{\gamma})\dot{\gamma}\label{eqn:stressexperiment}
\end{eqnarray}

Because the rheometer operates in a controlled stress mode, the above equations decouple. First we solve for (\ref{eqn:phiexperiment}) using the function \emph{ode15s}  provided with the \trademark{MATLAB} distribution. The function \emph{ode15s} performs well with stiff problems and its accuracy is recognized as medium to high. Equation (\ref{eqn:stressexperiment}) reduces to an algebraic equation for $\dot{\gamma}$ which we solve using the function \emph{fzero} provided with the \trademark{MATLAB} distribution as well. The function \emph{fzero}  uses a combination of bisection, secant, and inverse quadratic interpolation methods.

\begin{figure}
\begin{center}
(a)\scalebox{0.4}{\includegraphics{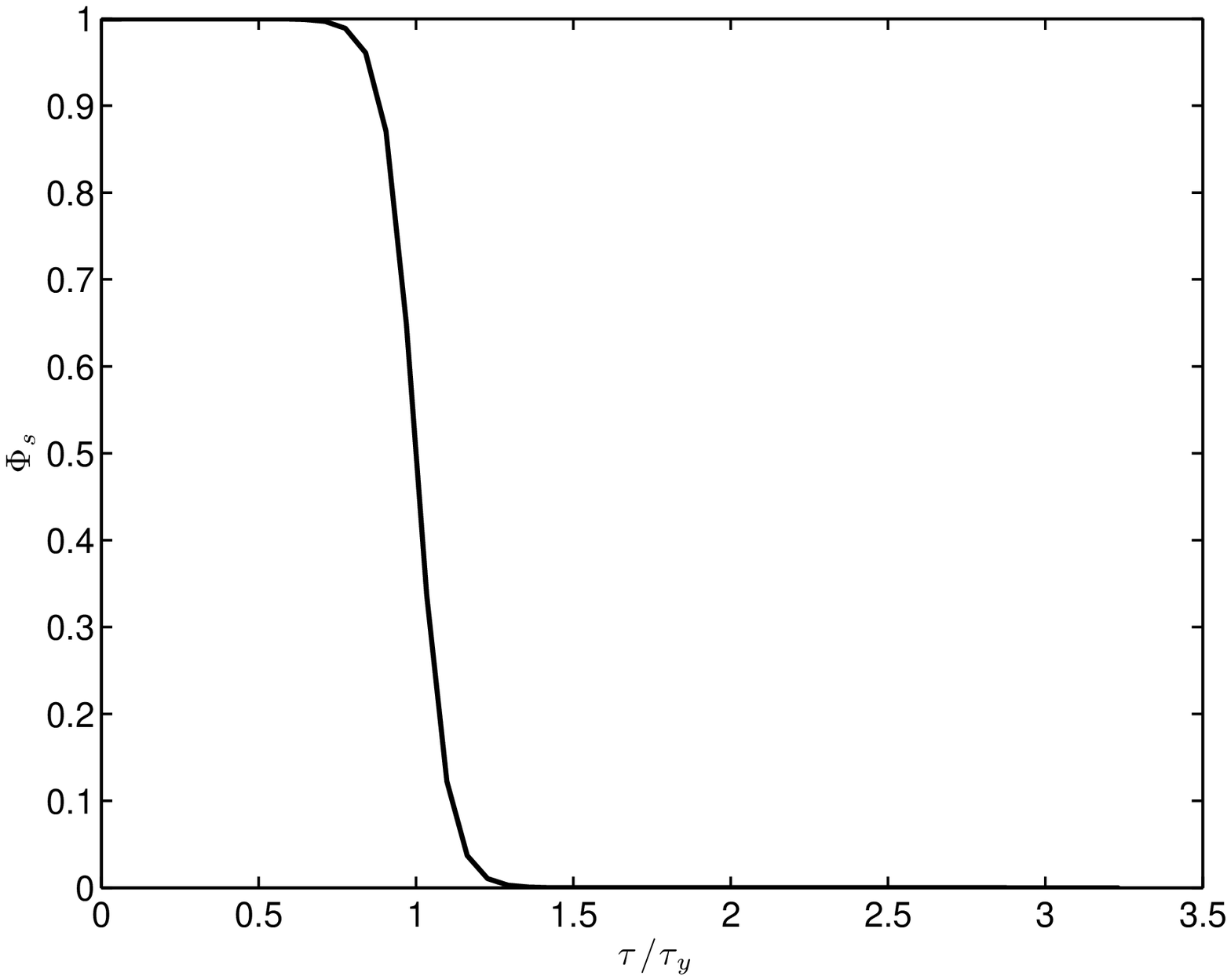}}
(b)\scalebox{0.4}{\includegraphics{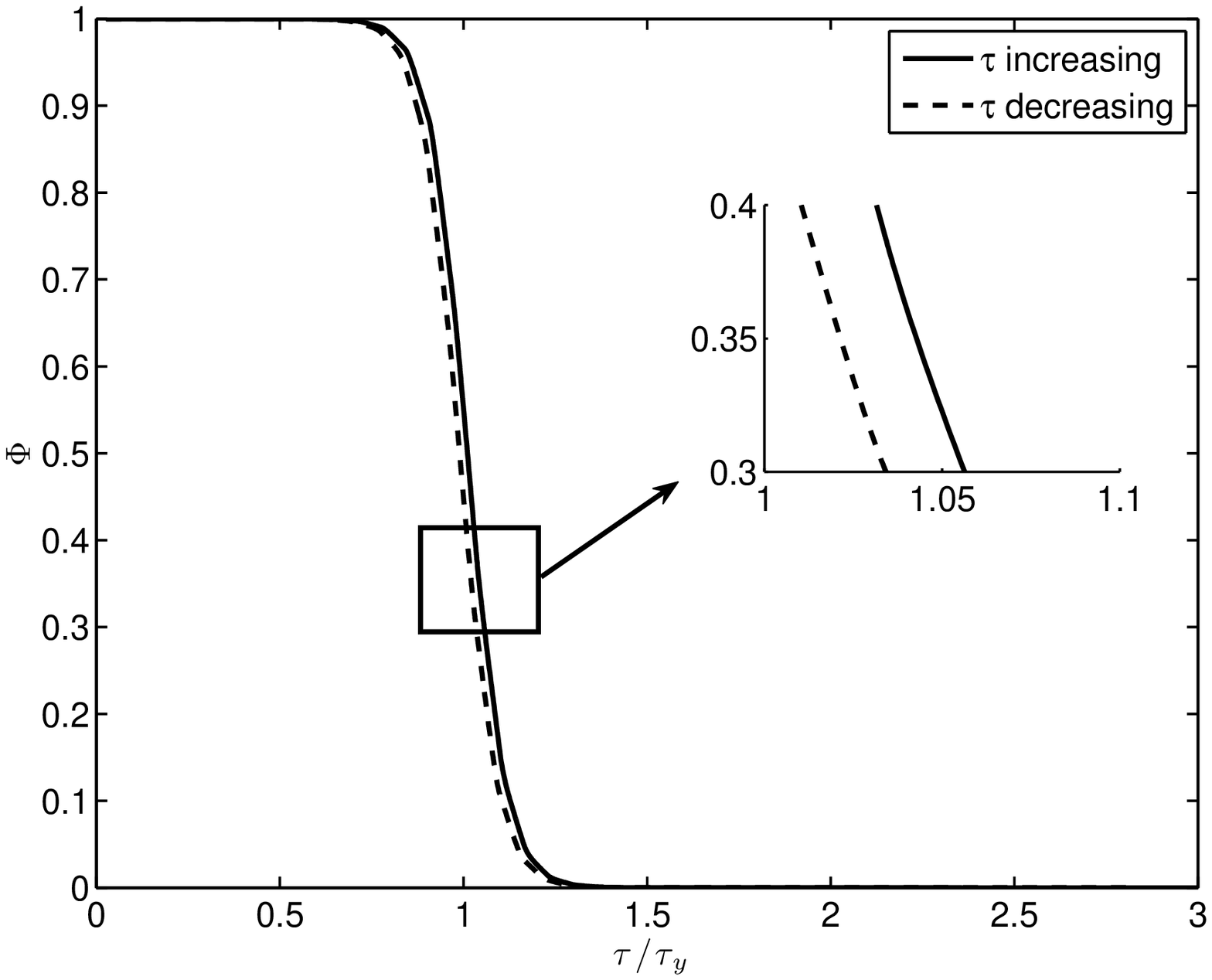}}
\end{center}
\caption{(a) Steady $\Phi$. (b) Transient $\Phi$. Both with $\tau_y=6.5$Pa $K_d=K_r=0.3$ and $w=0.1$.}
\label{fig:phi}
\end{figure}

\begin{figure}
\begin{center}
(a)\scalebox{0.4}{\includegraphics{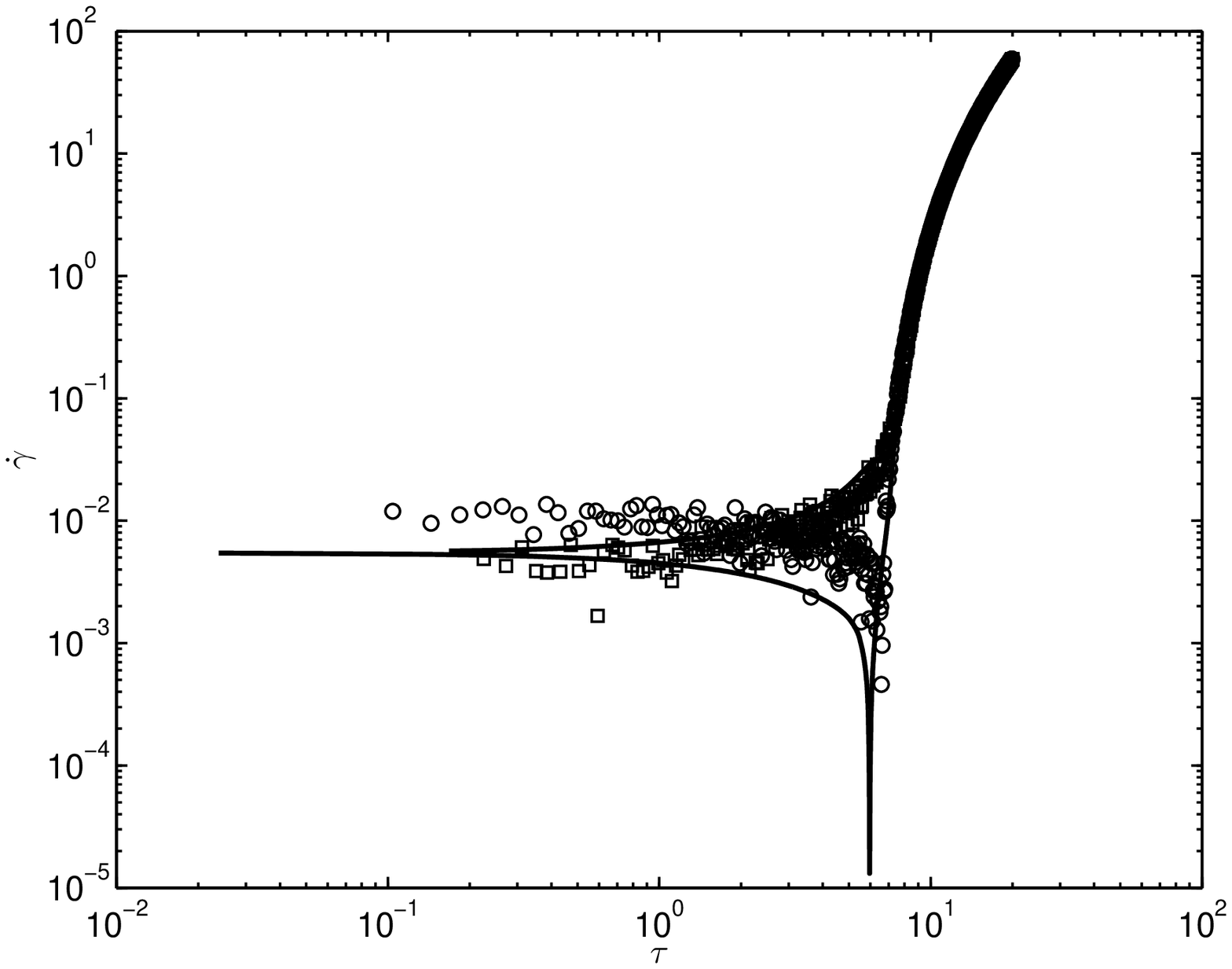}}
(b)\scalebox{0.4}{\includegraphics{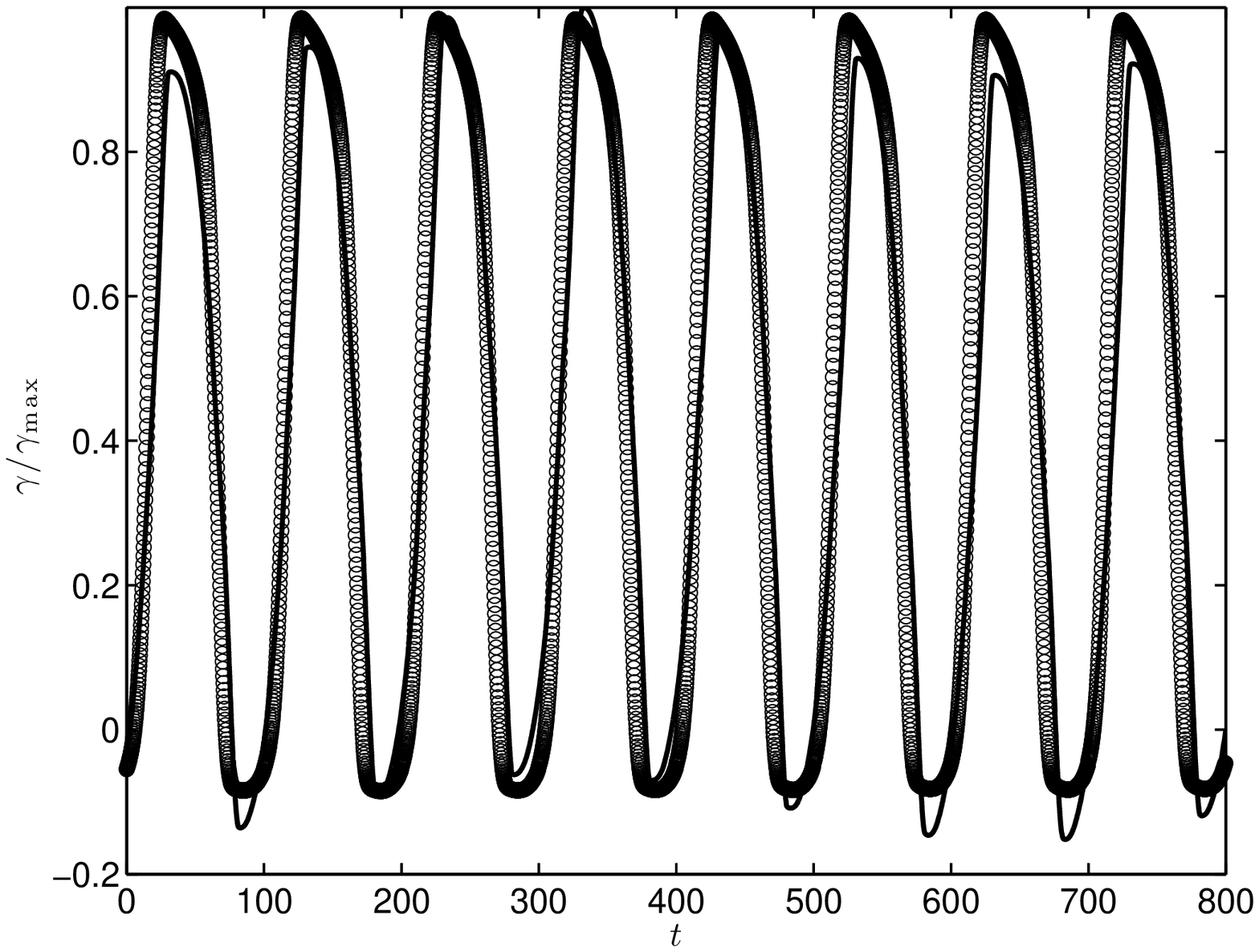}}
\end{center}
\caption{(a) Flow curve. (b) Strain time series. In bot plots we have $m=3$, $\mu_{\infty}=0.0102$, $\epsilon=0.01$, $\tau_y=6.5$, $K_d=K_r=0.3$ and $w=0.1$.}
\label{fig:experiment}
\end{figure}

In Figure \ref{fig:phi}a we present the steady state solution of equation (\ref{eqn:phiexperiment}) which reads
\begin{equation}
\Phi_s=\frac{f}{f+g}.
\end{equation}
In Figure \ref{fig:phi}b we present the solution of (\ref{eqn:phiexperiment}) using the same stress ramp as in the experiments presented in Ref. \cite{Burghelea09}. Clearly, for intermediate values of $\tau$, the decreasing stress branch of the solid concentration lags behind the increasing branch and a hysteresis is clearly visible. Having now both $\tau$ and $\Phi$ we can calculate $\dot{\gamma}$ from (\ref{eqn:stressexperiment}) using (\ref{Cassonvisc})  for $\mu(\dot{\gamma})$. The values for $m$, $\mu_{\infty}$ and $\epsilon$ are obtained from the fit performed in the rheometer. We obtain the elastic modulus $G$ from a linear fit of the experimentally measured dependence $\tau=G\gamma$ using the experimental data. The parameters  $K_d$, $K_r$ and  $w$ were chosen such that the numerical flow curve is in close agreement with the experimental flow curve. In Figure \ref{fig:experiment}a we compare the model discussed above with the experimental data. An excellent agreement is obtained, particularly at the point where $\dot{\gamma}=0$.

Finally,  we test the model against to oscillatory flow measurements where a harmonic forcing $\tau=\tau_0\sin(2\pi\kappa t)$ and the strain response $\gamma=\gamma(t)$ is monitored. This step is needed for a proper validation of the model, as the model involves three adjustable parameters ($K_d$, $K_r$ and  $w$). Note that $\dot{\gamma}=d\gamma/dt$. As before, having $\tau$ we solve for $\Phi$ using (\ref{eqn:phiexperiment}), then $\dot{\gamma}$ is found using (\ref{eqn:stressexperiment}), and finally we can use standard numerical integration to find $\gamma$. The results are presented in Figure \ref{fig:experiment}b.

One can conclude that using the values of the adjustable parameters that have validated the flow curves presented in Fig. \ref{fig:experiment}a a good agreement with the oscillatory measurements is obtained without any additional parametric adjustments, Fig. \ref{fig:experiment}b. We consider this result as a validation of the model initially proposed in \cite{Burghelea09} and revisited in Sec. \ref{sec_pmm} which further justifies employing it in the linear stability analysis presented through the rest of the paper.

%----------------------------------------------
\subsection{Non-dimensionalization}
To conclude this section we present the non-dimensional set of equations. Let
\[
\xb=\hat{\xb}L,~~~~\ub=\hat{\ub}U_{\max}, ~~~~t=\hat{t}L/U_{\max},~~~~ p=\hat{p}\rho U_{\max}^2, \taub=\hat{\taub}\mu U_{\max}/L
\]
where $\mu=\mu_s+\mu_{\infty}$ is the total viscosity of the completely  yielded fluid. Substituting all these into (\ref{eqnmotiondim})-(\ref{phieqndim}) we get:

\begin{eqnarray}
\frac{\partial \hat{\ub}}{\partial \hat{t}}+(\hat{\ub}\cdot{\bm \nabla})\hat{\ub}&=&-\nabla \hat{p}+\frac{\mu_r}{Re} {\bm \nabla} \cdot \bm{\hat{\gammabdot}}+\frac{1}{Re}{\bm \nabla} \cdot\hat{\taub} \label{eqnmotion}\\
{\bm \nabla} \cdot\hat{\ub}&=& 0 \label{masscons},\\
\hat{\taub}+We\hat{\mu}\left(\gammadot\right)\Phi \stackrel{\triangledown}{\hat{\taub}} &=& \hat{\mu}\left(\gammadot\right) \hat{\gammabdot}, \label{const}\\
\frac{\partial \Phi}{\partial \hat{t}}+(\hat{\ub}\cdot{\bm \nabla})\Phi&=&-\hat{g}(\hat{\tau})\Phi+\hat{f}(\hat{\tau})(1-\Phi).\label{phieqn}
\end{eqnarray}
where $\mu_r$ is the ratio of the solvent viscosity to the total viscosity, i.e. $\mu_s/\mu$. Finally:
\begin{eqnarray}
\hat{\mu}(\hat{\gammadot})&=&\left((1-\mu_r)^{1/m}+\left(\frac{B}{\sqrt{\hat{\gammadot}^2+\varepsilon^2}}\right)^{1/m}\right)^m,\label{Cassonviscndim}\\
\hat{g}(\hat{\tau})&=&\hat{K_d}\left(1+\tanh\left(\frac{\hat{\tau}-B}{\hat{w}}\right)\right),\label{geqnndim}\\
\hat{f}(\hat{\tau})&=&\hat{K_r}\left(1-\tanh\left(\frac{\hat{\tau}-B}{\hat{w}}\right)\right),\label{feqnndim}
\end{eqnarray}
where
\[
K_d=\hat{K_d}L/U_{\max},~~~~K_r=\hat{K_r}L/U_{\max},~~~~w=\hat{w}\mu U_{\max}/L.
\]
The Reynolds ($Re$), Weissenberg ($We$) and Bingham ($B$) numbers are defined:
\[
Re=\frac{\rho L U_{\max}}{\mu},~~~~We=\frac{\lambda_H U_{\max}}{L},~~~~B=\frac{\tau_y L}{U_{\max}\mu}
\]
with the relaxation time $\lambda_H=\mu_{\infty}/G$.

Through the rest of the paper only non-dimensional variables will be used and the ``hat'' shall be dropped for simplicity.

%%%%%%%%%%%%%%%%%%%%%%%%%%%%%%%%%%%%%%%%%%%%%%%%%%%%%%%%%%%%%%%%%%%%%%%%
%%%%%%%%%%%%%%%%			`			Linear stability analysis reg. model				%%%%%%%%%%%%%%%%
%%%%%%%%%%%%%%%%%%%%%%%%%%%%%%%%%%%%%%%%%%%%%%%%%%%%%%%%%%%%%%%%%%%%%%%%
\section{Linear stability analysis for shear-thinning regularized Casson's Model}

First we consider the linear stability of the regularized Casson model. We note that our main goal is to investigate the effect the solid and solid-fluid regions have on the stability of the flow. It is known that a viscoplastic fluid is stable to all infinitesimal perturbations (at least no instabilities have been found numerically) see \cite{Nouar07b}, therefore a natural second step towards our goal is to ask ourselves how the presence of a highly stratified viscosity affects the stability of the flow, thus we consider the case $\Phi\equiv 0$. Linear stability analysis for a regularized Bingham model was carried out by Frigaard and Nouar in \cite{Frig05}. There they showed that as $\epsilon\rightarrow 0$ the velocity field using a regularized viscosity converges to the velocity field using the non-regularized viscoplastic viscosity. While considering the problem of linear stability they showed that the spectrum for the regularized problem had some physical spurious eigenvalues, not present in the Bingham problem. The stability of the eigenvalues depended on $\epsilon$, we will discuss further about this in Sec. \ref{subsec:validationmodOrr-Som}. Here we are not interested in the distinguished limit $\epsilon\rightarrow 0$. Based on experimental data we are aware of that typically  $O(10^{-3})\leq \epsilon \leq O(10^{-2})$, therefore for the rest of the paper we fix $\epsilon=0.002$ unless otherwise stated. 

For the rest of this section we set $\Phi=0$, $\mu_s=0$ and  $\mu=\mu_{\infty}$, then the equation of motion reads
\begin{equation}\label{eq:regmomentum}
\frac{D\ub}{Dt} = -\grad{p} +\frac{1}{Re} \grad\cdot\taub,
\end{equation}
along with the incompressibility condition $\grad\cdot\ub=0$. 

The constitutive equation reduces to:
\[
\taub = \mu(\gammadot)\gammabdot,
\]

with $\mu(\gammadot)$ satisfying (\ref{Cassonviscndim}) with $\mu_r=0$.

%-------------------------------------
\subsection{Baseflow profile}\label{baseflowreg}

We consider shear flows of the form $\Ub=(U(y),0,0)$ with $y\in[-1,1]$, then (\ref{eqnmotion}) reduces to:

\begin{equation}
\frac{d\tau_{xy}}{dy}=Re\frac{dp}{dx}\label{eqn:baseflowreg}
\end{equation}

with the usual non-slip boundary conditions $U(-1)=U(1)=0$ at the walls. It is clear that in this situation $\gammadot=|dU/dy|$ and because $U(y)$ reaches its maximum at the center of the channel we have: $\gammadot=-dU/dy$ if $y\in [0,1]$ and $\gammadot=dU/dy$ otherwise. Having this in mind and integrating once (\ref{eqn:baseflowreg}) we have the following algebraic equation for $\gammadot$:

\begin{equation}\label{eq:baseflowgamma}
|y|\frac{dp}{dx}Re + \mu(\gammadot)\gammadot = 0.
\end{equation}

We solve this non-linear equation for $\gammadot$ using \emph{fzero} in MATLAB for each point of our discrete domain. Once we have the approximation for $\gammadot$ we integrate it with respect to $y$ and use our boundary conditions to get $U(y)$.  We feed this result to an user-built function in MATLAB and use \emph{fzero} again to find $Re(dp/dx)$ such that $\max_y(U(y))=1$.

\begin{figure}
\begin{center}
\scalebox{0.5}{\includegraphics{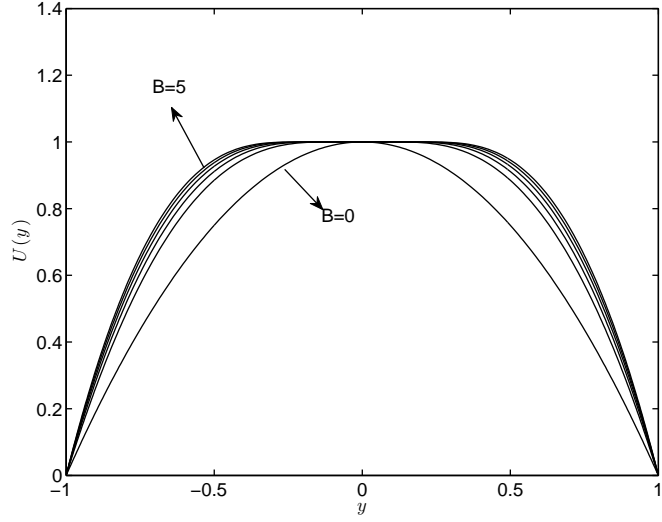}}
\end{center}
\caption{Base flow profile for the regularized shear-thinning Casson model with $\epsilon=0.002$, $m=2.5$ and $B=0,~1,~2,~3,~4,~5$.}
\label{fig:baseflowreg}
\end{figure}

We present different examples of the base flow for different Bingham numbers in Figure \ref{fig:baseflowreg}, clearly as we increase B we can see the existence of a pseudo-plug region around the center line of the channel. Recall that this is not a true plug, i.e. the velocity there is not constant. 

%---------------------------------------
\subsection{Linearised momentum equation and tangent viscosity}
As is common in linear stability analysis we consider an infinitesimal perturbation $(\varepsilon\ub',\varepsilon p')$ superimposed upon the base flow and linearize  the momentum equation (\ref{eq:regmomentum}) around $(\Ub,p)$ to get:

\begin{equation}\label{eq:linearised-momentum-equation}
\ddy{\ub'}{t}+(\ub'\cdot\gradb)\Ub+(\Ub\cdot\gradb)\ub' = -\grad{p'} + {\bm \grad}\cdot \taub',
\end{equation}
where $\taub^{\prime}$ is the stress perturbation given by:
\begin{equation}
\taub^{\prime}= \mu(\Ub)\gammabdot(\ub')+\tilde{\mu}\gammabdot(\Ub).\label{eqn:regstresspert}
\end{equation}
where $\tilde{\mu}$ is the viscosity perturbation and it is given by:
\begin{equation}
\tilde{\mu}=\gammadot_{ij}(\ub')\ddy{\mu}{\gammadot_{ij}}(\Ub).
\end{equation}

\begin{figure}
\begin{center}
(a)\scalebox{0.5}{\includegraphics{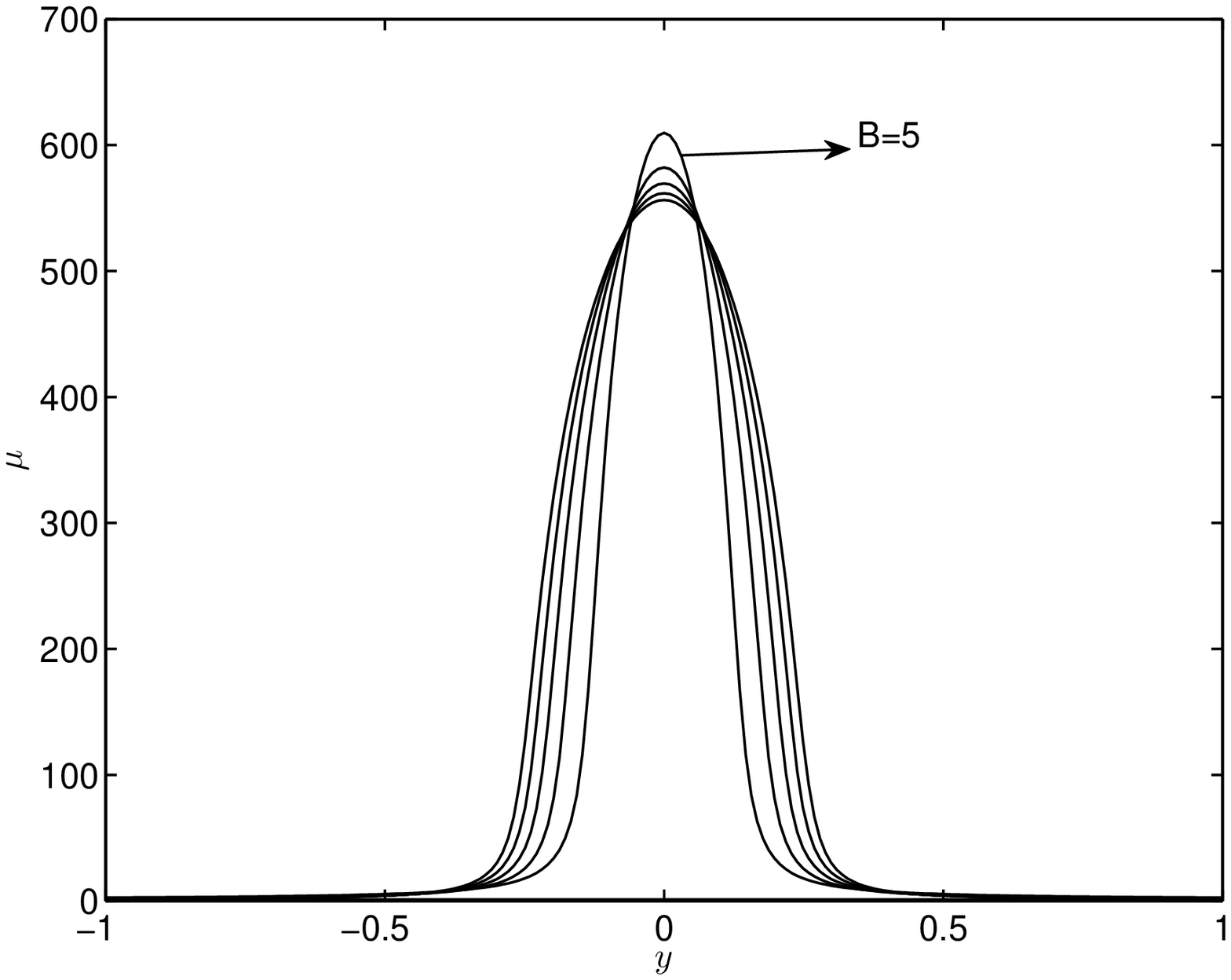}}
(b)\scalebox{0.5}{\includegraphics{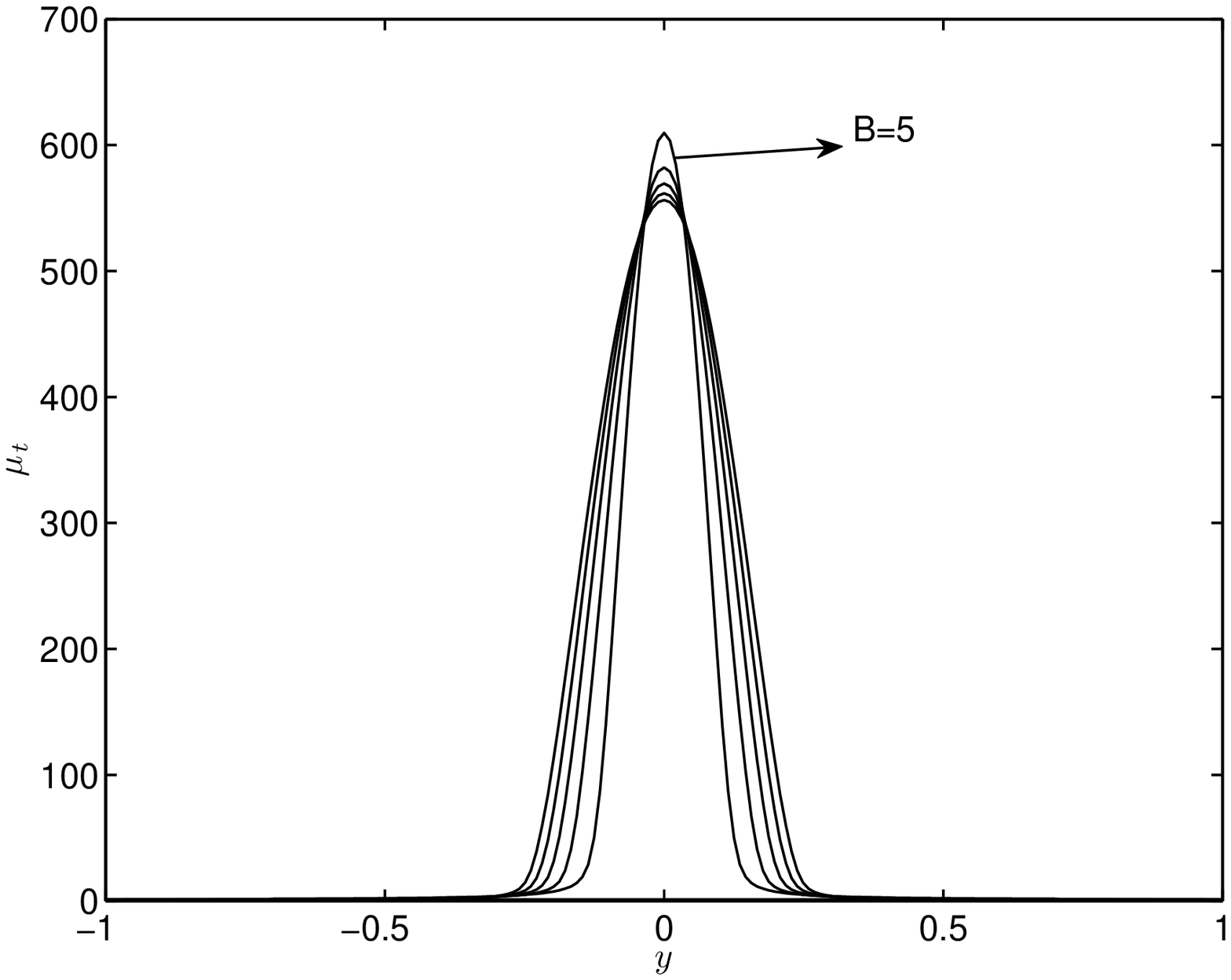}}
\end{center}
\caption{(a) Non-linear viscosity $\mu(y)$ and (b) tangent viscosity $\mu_t(y)$ with $\epsilon=0.002$, $m=2.5$ and $B=0,~1,~2,~3,~4,~5$.}
\label{fig:muandmutreg}
\end{figure}

Given the fact that the flow we consider is unidirectional, it can be shown that (see \cite{Nouar07}):

\begin{equation}
\tau_{ij} = \left\{
\begin{array}{lcl}
\mu(\Ub)\gammadot_{ij}(\ub') & \mathrm{for} & ij \neq xy,yx\\
\mu_{t}(\Ub)\gammadot_{ij}(\ub') & \mathrm{for} & ij = xy,yx,\\
\end{array}
\right.
\end{equation}
where
\begin{equation}\label{def:tangent-viscosity}
\mu_{t} = \mu(\Ub) + \frac{d\mu}{d\gammadot_{xy}}(\Ub)\gammadot_{xy}(\Ub)
\end{equation}
is the tangent viscosity. For a one-dimensional shear flow, the tangent viscosity is defined by $\mu_{t}=d\tau_{xy}/d\gammadot_{xy}$ whereas the effective viscosity is defined as $\mu=\tau_{xy}/\gammadot_{xy}$. In Figure \ref{fig:muandmutreg} we show some examples of the non-linear viscosity $\mu(y)$ and $\mu_t(y)$ for increasing Bingham numbers, note that the functions are smooth, $\mu\geq\mu_t$ and that $\max_y\mu(y)\equiv\max_y\mu_t(y)\sim B/\epsilon$.
 For a detailed description of the tangent viscosity concept we refer the reader to \cite{Nouar07}.

%------------------------------------
\subsection{Modified Orr-Sommerfeld and Squire equations}

Writing equation (\ref{eq:linearised-momentum-equation}) in terms of the normal velocity $v'(\xb,t)$ and the normal vorticity $\eta(\xb,t)=\dy{u'}/\dy{z} - \dy{w'}/\dy{x}$ and assuming modal solutions of the form:
\begin{equation*}\begin{aligned}
&v'(\xb,t) = \hat{v}(y)\exp[i(\alpha{x}+\beta{z}-\omega{t})],\\
&\eta(\xb,t) = \hat{\eta}(y)\exp[i(\alpha{x}+\beta{z}-\omega{t})],
\end{aligned}\end{equation*}

we get the following eigenvalue problem for the frequency $\omega$:
\begin{equation}\label{eq:modified-os-sq}
\left(\begin{array}{cc}
\mathcal{L} & \mathcal{C}_{1}\\
\mathcal{C}_{2} & \mathcal{S}
\end{array}\right)
\left(\begin{array}{c}
v'\\
\eta
\end{array}\right) =
\omega \left(\begin{array}{c}
\tilde{\lapl}v'\\
\eta
\end{array}\right),
\end{equation}
where
\begin{eqnarray}
\mathcal{L} &=& \alpha[U\tilde{\lapl}-(D^{2}U)] + \frac{i}{Re}[\mu\tilde{\lapl}^{2}+2(D\mu)D^{3}+(D^{2}\mu)D^{2}-2k^{2}(D\mu)D+k^{2}(D^{2}\mu)]\nonumber\\
& +&\frac{i\alpha^{2}}{Re\ k^{2}}(D^{2}+k^{2})[(\mu_{t}-\mu)(D^{2}+k^{2})],\label{modOrr-Somop}\\
\mathcal{C}_{1}&=&-\frac{i\alpha\beta}{Re\ k^{2}}(D^{2}+k^{2})[(\mu_{t}-\mu)D],\\
\mathcal{C}_{2}&=&\beta(DU)-\frac{i\alpha\beta}{Re\ k^{2}}D[(\mu_{t}-\mu)(D^{2}+k^{2})],\\
\mathcal{S}&=&\alpha{U} +\frac{i}{Re}\mu\tilde{\lapl}+\frac{i}{Re}(D\mu)D + \frac{i}{Re}\frac{\beta^{2}}{k^{2}}D[(\mu_{t}-\mu)D],\label{modSquireop}
\end{eqnarray}
with $k^{2}=\alpha^{2}+\beta^{2}$, $D=d/dy$, $\tilde{\lapl}=D^{2}-k^{2}$. Together with the boundary conditions 
\begin{equation}
v=Dv=\eta=0~~\mbox{at}~~ y=\pm1.\label{regbc}
\end{equation}

%------------------------------------
\subsection{Bounds for the Squire and Orr-Sommerfeld modes and one-dimensional stability}

In this section we present bounds for the eigenvalues of the modified Orr-Sommerfeld and the modified Squire operators, equations (\ref{modOrr-Somop}) and (\ref{modSquireop}). We follow closely the work by Joseph \cite{Joseph68} and  Davies and Reid \cite{DavisReid77}. For the rest of the section instead of working with the frequency $\omega$ as the eigenvalue we consider the wave speed, $c=\omega / \alpha$ unless otherwise stated.

As in \cite{Joseph68} we define 
\[I_{n}^{2}=\int^{1}_{-1}|v^{(n)}|^{2}\dd y\]
and take $v\in\Hhat$, where $\Hhat$ is a complex-valued Hilbert space completed under the norm $I_2^2$ by the addition of limit points of sequences of functions in $C^4([-1,1])$ satisfying (\ref{regbc}). The following isoperimetric inequalities hold in $\Hhat$:
\begin{lem}\label{thm:poincare}
Let $v\in\overline{\mathcal{H}}$, then
\begin{equation*}\begin{aligned}
I_{1}^{2}\geq\lambda_{1}^{2}I_{0}^{2},\ \ \ \ \ I_{2}^{2}\geq\lambda_{2}^{2}I_{1}^{2},\ \ \ \ \ I_{2}^{2}\geq\lambda_{3}^{2}I_{0}^{2}\\
\lambda_{1}^{2}=\frac{\pi^{2}}{4},\ \ \ \ \ \lambda_{2}^{2}=\pi^{2},\ \ \ \ \ \lambda_{3}^{2}=(2.365)^{4},
\end{aligned}\end{equation*}
where $\lambda_3$ is the smallest eigenvalue of a vibrating rod with displacement $v$ satisfying (\ref{regbc}).
\end{lem}

\subsubsection{Bounds for the Squire and Orr-Sommerfeld modes}

\begin{thm}\label{thm:nn-squire}
\emph{(Damped Squire modes)} Let $c(\alpha,\beta,Re)$ be any eigenvalue of the homogeneous modified Squire's equation 
\begin{equation*}
\frac{1}{\alpha}\mathcal{S}\eta=c\ \eta, \ \ \ \ \ \ \eta(\pm1)=0.
\end{equation*}
Then
\begin{equation}\begin{aligned}
U_{min}<&c_{r}<U_{max}\\
&c_{i}<-p\frac{(\pi^{2}/4)+k^{2}}{\alpha Re}
\end{aligned}\end{equation}
where $p=\min\{\mu_{t}(y)|-1\leq{y}\leq1\}$.
\end{thm}

See \ref{App:Squire-proof} for the proof.

\begin{thm}\label{thm:nn-joseph}
Let $c(\alpha,Re)$ be any eigenvalue of the modified Orr-Sommerfeld equation
\begin{equation}\label{thm:nn-joseph-os}
\frac{1}{\alpha}\mathcal{L}v = c(D^{2}-\alpha^{2})v\ \ \ \ \ \ \ \ v(\pm1)=v^{\prime}(\pm1) = 0.
\end{equation}
Let $q=\max\{|U^{\prime}(y)| : -1\leq{y}\leq1\}$ and $p=\min\{\mu_{t}(y)|-1\leq{y}\leq1\}$, then we have the following results:
\begin{enumerate}
\item \begin{equation}\label{thm:nn-joseph1}
c_{i}\leq\frac{q}{2\alpha} - \frac{p}{\alpha Re}\left(\frac{\pi^{2}(\pi^{2}+\alpha^{2})}{\pi^{2}+4\alpha^{2}}+\alpha^{2}\right).
\end{equation}
\item No amplified disturbances (modes with $c_{i}>0$) of \eqref{thm:nn-joseph-os} exist if
\begin{equation}\begin{aligned}\label{thm:nn-joseph2}
q\alpha{Re} &< p\cdot{}f(\alpha) \equiv \max[M_{1}, M_{2}],\\
M_{1} &= \lambda_{3}\pi + 2^{3/2}\alpha^{3}\\
M_{2} &= \lambda_{3}\pi + \pi\alpha^{2},\\
\end{aligned}\end{equation}
where $\lambda_{3}=(2.356)^{2}$.
\item \begin{equation}\begin{aligned}\label{thm:nn-joseph3}
&U_{min}^{\prime\prime}\leq0: &U_{min}&<c_{r}<U_{max}+\frac{2U_{max}^{\prime\prime}}{\pi^{2}+4\alpha^{2}}\\
&U_{min}^{\prime\prime}\leq0\leq{}U_{max}: &U_{min}+\frac{2U_{min}^{\prime\prime}}{\pi^{2}+4\alpha^{2}}&<c_{r}<U_{max}+\frac{2U_{max}^{\prime\prime}}{\pi^{2}+4\alpha^{2}}\\
&U_{max}^{\prime\prime}\leq0: &U_{min}+\frac{2U_{min}^{\prime\prime}}{\pi^{2}+4\alpha^{2}}&<c_{r}<U_{max}.\\
\end{aligned}\end{equation}
\end{enumerate}
\end{thm}
See \ref{App:Orr-Somm-proof} for a proof.

We should note that these results are not restrictive to regularized viscoplastic fluids, they hold for any shear-thinning (we are using the fact that $\mu$ is a decreasing function of $\gammadot$) viscosity function.

\subsubsection{One-dimensional stability}

From Theorem \ref{thm:nn-squire} we have that the Squire modes are always damped, thus for the rest of this section we consider only the modified Orr-Sommerfeld equation and $\alpha=0$. If instead of working with the perturbation of the normal velocity $v$ we consider a stream function formulation, i.e. $u=\Psi_y$ and $v=-\Psi_x$, and taking a modal solution of the form $\Psi=\phi(y)\exp[i(\alpha{x}-\omega{t})]$, then we can interchange $v$ for $\phi$  in (\ref{eq:modified-os-sq}). Note that $\phi\in\Hhat$ and Theorem \ref{thm:nn-joseph} holds for this formulation. Therefore using (\ref{pf:nn-joseph4}) we have a bound for the imaginary part of the frequency $\omega$:

\begin{equation}\label{omegaregbound}
\omega_{i}<\frac{q\alpha I_{1}I_{0}}{I_{1}^{2}+\alpha^{2}I_{0}^{2}} -\frac{({Re})^{-1}p(I_{2}^{2}+2\alpha^{2}I_{1}^{2}+\alpha^{4}I_{0}^{2}))}{I_{1}^{2}+\alpha^{2}I_{0}^{2}}.
\end{equation}

Here we are interested in one-dimensional perturbations, i.e. $\alpha=0$. This is a special case in terms of boundary conditions. The fact that $v\equiv 0$ does not mean that $\phi=0$ at $y=\pm 1$.  This boundary condition can be obtained if the perturbation to the flow rate $\int_{-1}^1u(y)\dd y$ is taken to be zero and a uniform pressure gradient in the $x$-direction is allowed, see \cite{Renardy86}. Now, even for the case $\alpha=0$ we have $\phi\in\Hhat$ and (\ref{omegaregbound}) becomes:
\begin{equation}\label{omegaregonedim}
\omega_{i}\leq - \frac{p\pi^2}{Re},
\end{equation}
where we used the isoperimetric inequalities defined in Lemma \ref{thm:poincare}.
Therefore, one-dimensional perturbations are always linearly stable. Note again that this result is valid for any decreasing function $\mu(\gammadot)$.

%---------------------------------------------------

\subsubsection{Solution of modified Orr-Sommerfeld equation and code validation}\label{subsec:validationmodOrr-Som}

There is no equivalent of Squire's theorem for generalized Newtonian fluids but some authors have performed  several numerical tests for large range of axial and transverse wave numbers ($\alpha$ and $\beta$ respectively) for different non-linear viscosity functions $\mu (\gammadot)$, see \cite{Nouar07,Nouar07b,Nouar09,Nouar10}. Their numerical results show that the lowest critical Reynolds number is obtained for spanwise homogeneous perturbations ($\beta=0$, $\alpha\neq0$). We should also note that in the situation of a homogeneous streamwise perturbation ($\alpha=0$, $\beta\neq0$), the imaginary part of $\omega$ satisfies:

\begin{equation}
\omega_i=-\frac{1}{Re}\frac{\int_{-1}^1\mu\left(4\beta^2|Dv|^2+|D^2v+\beta^2v|^2\right)\dd y}{\int_{-1}^1|Dv|^2+\beta^2|v|^2\dd y}<0,
\end{equation}
which comes from letting $\alpha=0$, multiplying by the complex conjugate of $v$, say $v^*$, integrating by parts and taking imaginary part in (\ref{eq:modified-os-sq}). Note that if $\alpha=0$ then $\mathcal{L}$ and $\mathcal{S}$ decouple, thus the flow is unconditionally stable.  Using these facts we would only consider 2D perturbations in the streamwise direction for the rest of the section. We know from Theorem \ref{thm:nn-squire} that all the Squire modes are always damped, thus we are concerned only with the modified Orr-Sommerfeld equation:

\begin{equation}\label{Orr-Somreg}\begin{aligned}
\alpha c (D^2-\alpha^2)v =& \alpha[U(D^2-\alpha^2)-D^{2}U] v- \frac{4i\alpha^2}{Re}D(\mu D)v\\
& +\frac{i}{Re}(D^{2}+\alpha^{2})[\mu_{t}(D^{2}+\alpha^{2})]v,\\
\end{aligned}\end{equation}
Together with the boundary conditions
\begin{equation}
v=Dv=0~~\mbox{at}~~ y=\pm1.\label{regbcorr}
\end{equation} 

\begin{figure}
\begin{center}
(a)\scalebox{0.4}{\includegraphics{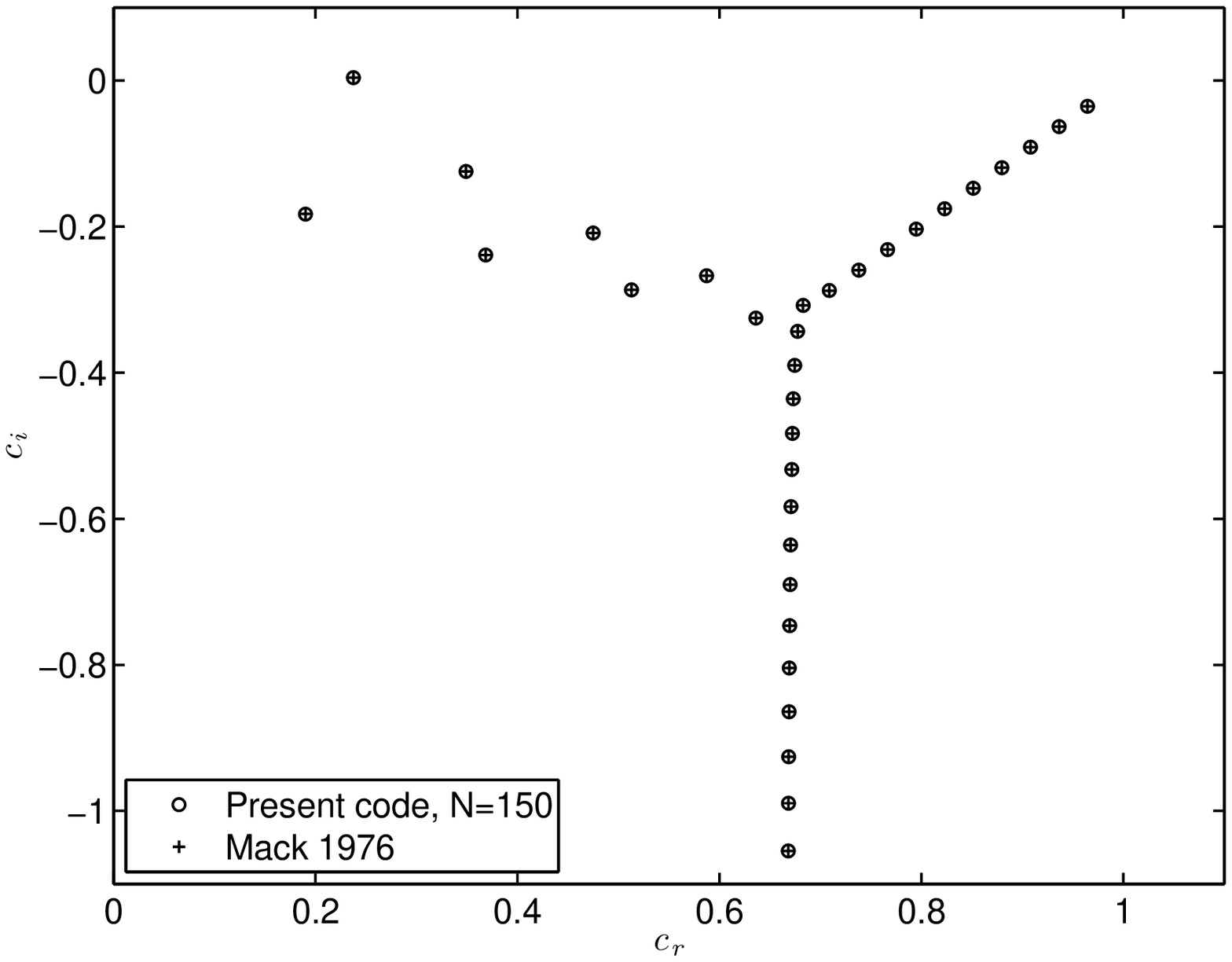}}\\
(b)\scalebox{0.4}{\includegraphics{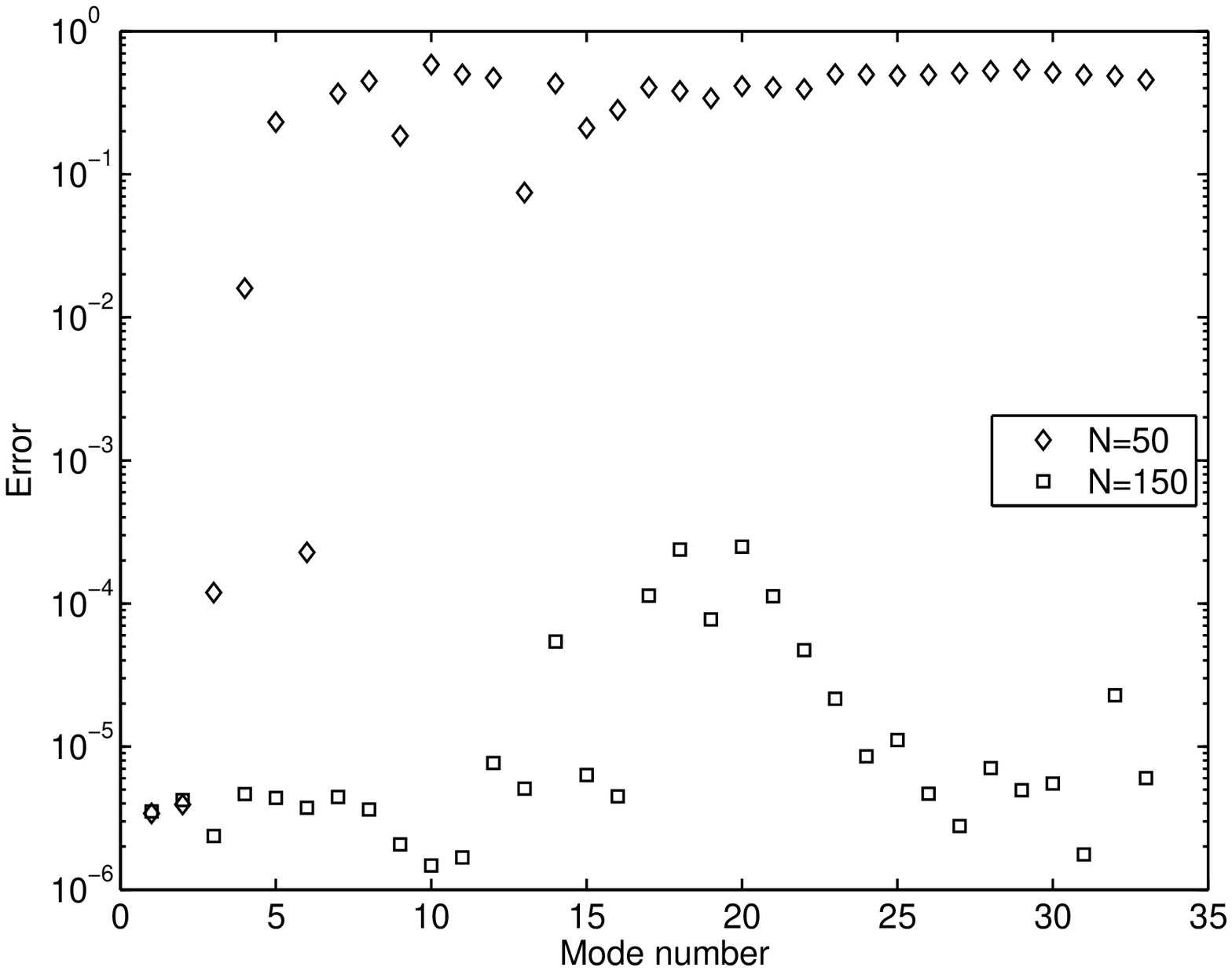}}
\end{center}
\caption{(a) Spectrum for Newtonian case, present code with $N=150$ (o), Mack \cite{Mack76} ($+$). (b) Error norm of present code with $N=50$ (diamonds) and $N=150$ (squares) vs Mack \cite{Mack76}. All for $Re=10000$ and $\alpha=1$.}
\label{fig:validationnewt}
\end{figure}

\begin{figure}
\begin{center}
\scalebox{0.5}{\includegraphics{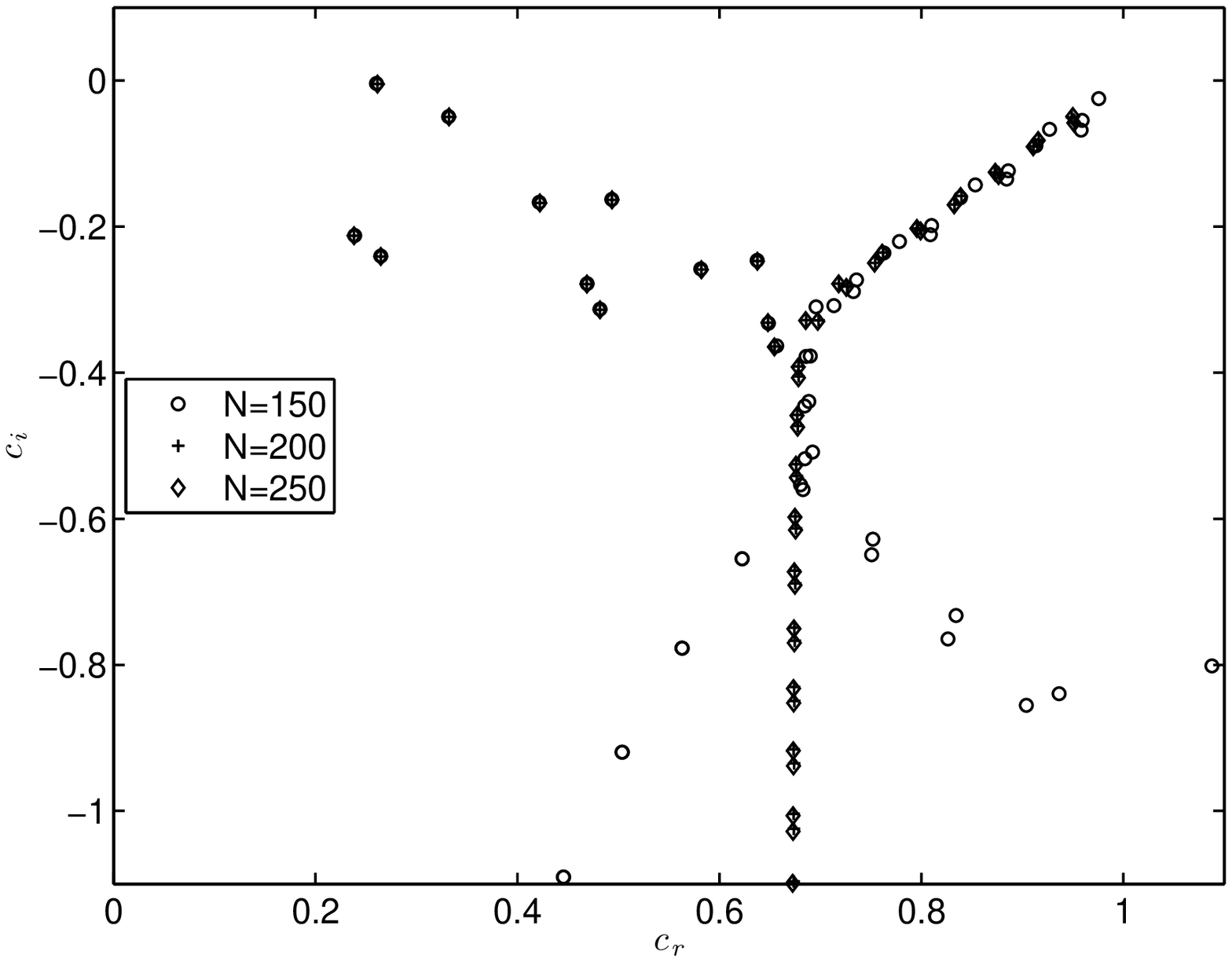}}
\end{center}
\caption{Spectrum for regularized shear-thinning Casson's model, $N=150$ (o), $N=200$ ($+$), and $N=250$($\Diamond$). For $Re=10000$ and $\alpha=1$. All calculations with $\epsilon=0.002$.}
\label{fig:validationreg}
\end{figure}

We solve the eigenvalue problem (\ref{Orr-Somreg})-(\ref{regbcorr}) by discretizing using Chebyshev polynomials in the usual fashion, as described for example in \cite{Schmid01}. In order to validate our code, we benchmark by solving the Newtonian problem $m=1$ and $B=0$, i.e. $\mu\equiv\mu_t\equiv 1$ in (\ref{Orr-Somreg}) and compare with results obtained by Mack \cite{Mack76}.  In Figure \ref{fig:validationnewt}a we present the first 33 modes calculated using our code with $N=150$ and the ones reported by Mack with the eigenvalues of the two computations overlaying one another. Figure \ref{fig:validationnewt}b shows the error norm between our calculations and Mack's for $N=50$ and $N=150$. As expected the least stable eigenvalue has converged but the split of the $S$-branch is present when $N$ is not large enough.  
In order to stably compute the spectrum of (\ref{Orr-Somreg})-(\ref{regbcorr}) we need substantially more nodes than for the Newtonian problem. In Figure \ref{fig:validationreg} we present the calculations using $N=150,~200,~250$. The first thing to note is that even with $N=150$ the S-branch of the spectrum has not converged. Another important fact about the spectrum is the presence of an R-branch, which was first documented by Frigaard and Nouar in \cite{Frig05}. They found that this branch is physically spurious, meaning that these eigenvalues do not exist for the Orr-Sommerfeld operator corresponding to the viscoplastic model and their stability depends on $\epsilon$. As $\epsilon\rightarrow 0$ we can find unstable modes from this R-branch even for moderate Reynolds numbers. We have found that by choosing a smooth regularized non-linear viscosity (either the one due to Becovier et. al. or the one due to Papanastasiu) increases the range of $\epsilon$ for which the R-branch remains stable and the least stable eigenvalue belongs to the A-branch as expected. We have fixed $\epsilon=0.002$ for all our results and we have checked that the critical eigenvalue is indeed in the A-branch. For the rest of the section we fix the number of nodes to be $N=250$.

%----------------------------------------------------

\subsection{Numerical results}

\begin{figure}
\begin{center}
(a)\scalebox{0.5}{\includegraphics{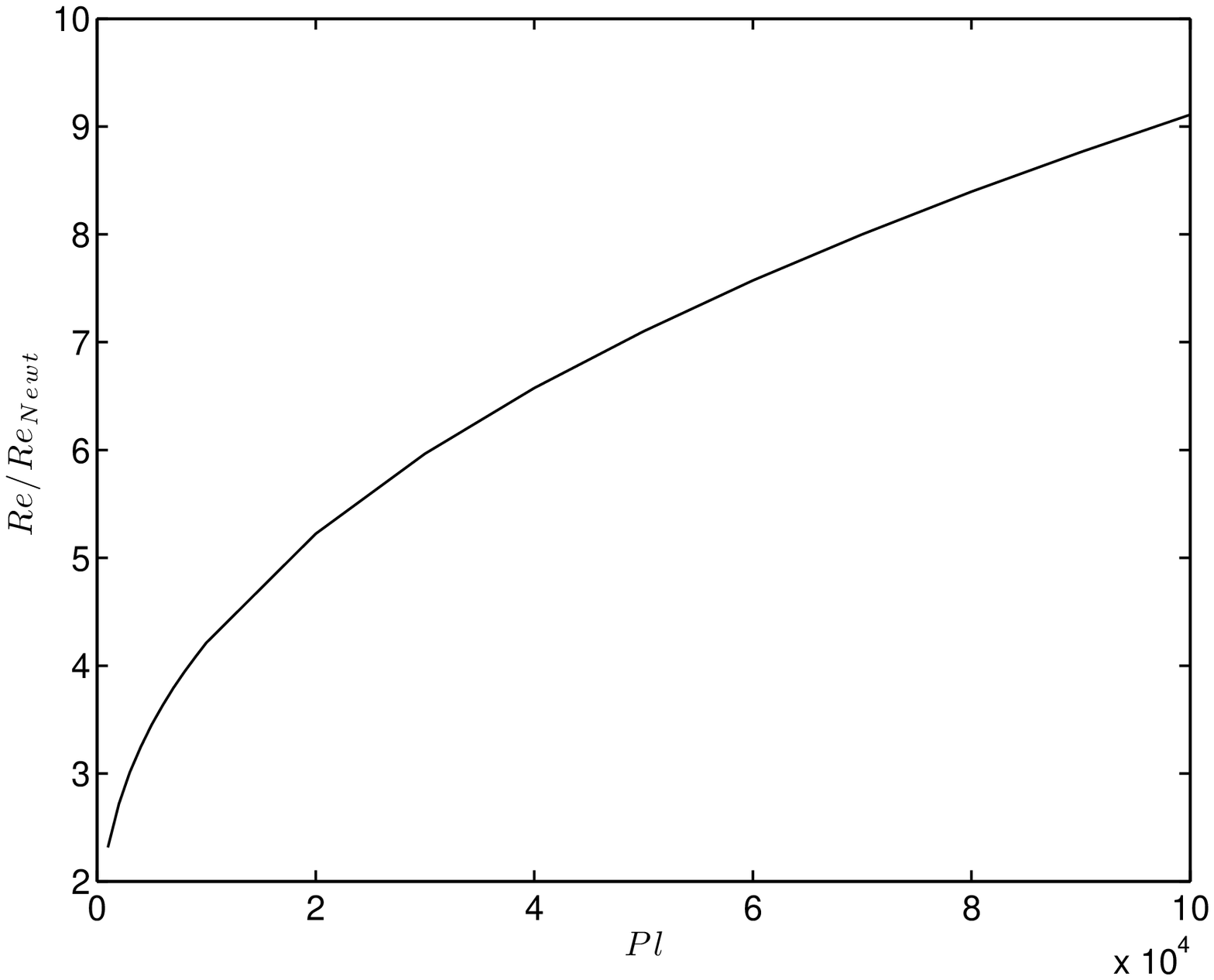}}
(b)\scalebox{0.5}{\includegraphics{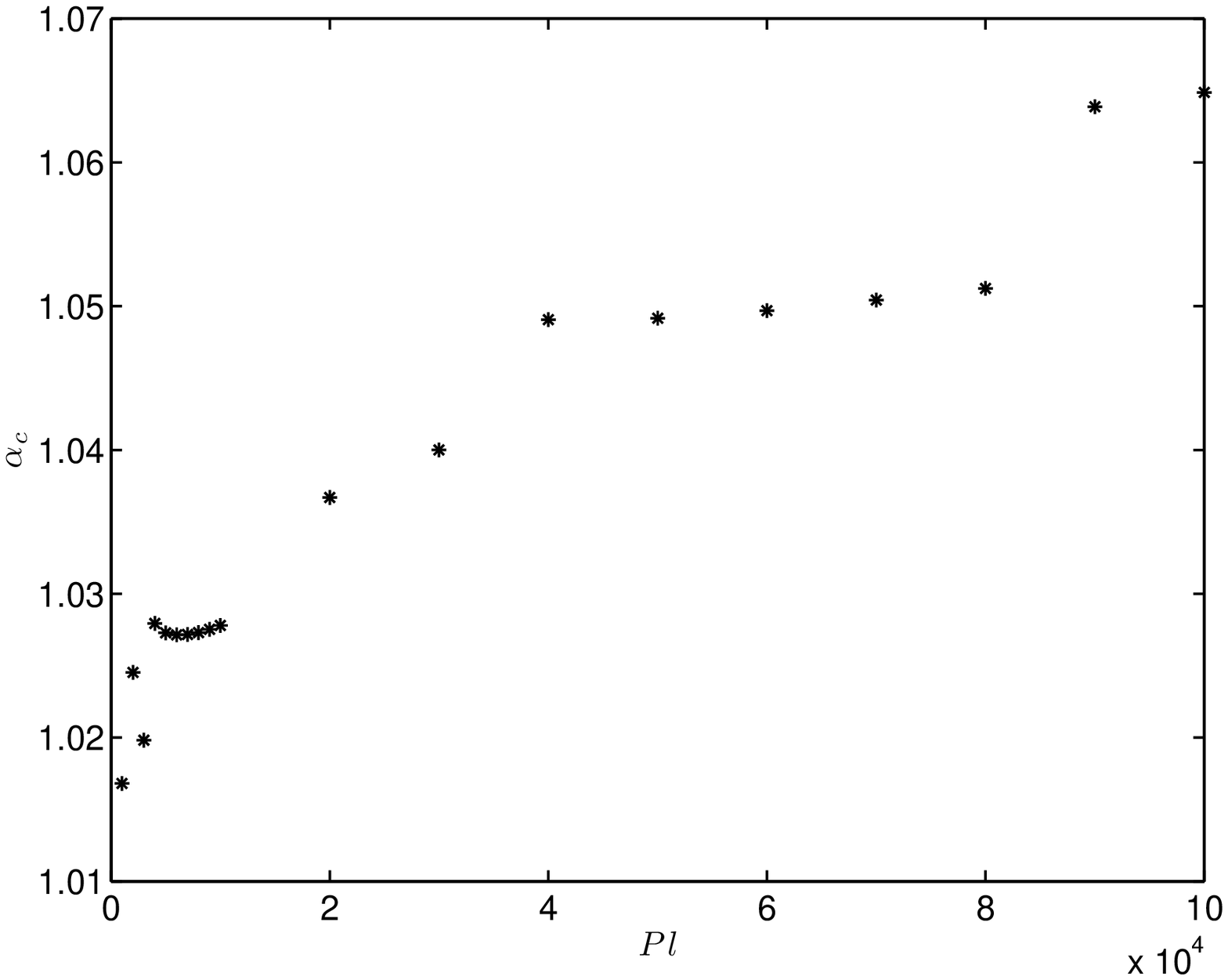}}
\end{center}
\caption{(a) Normalized  critical $Re$ for increasing $Pl$ and $m=2.5$. (b) Critical wave number . All calculations with $\epsilon=0.002$.}
\label{fig:resultsregpl}
\end{figure}

\begin{figure}
\begin{center}
(a)\scalebox{0.5}{\includegraphics{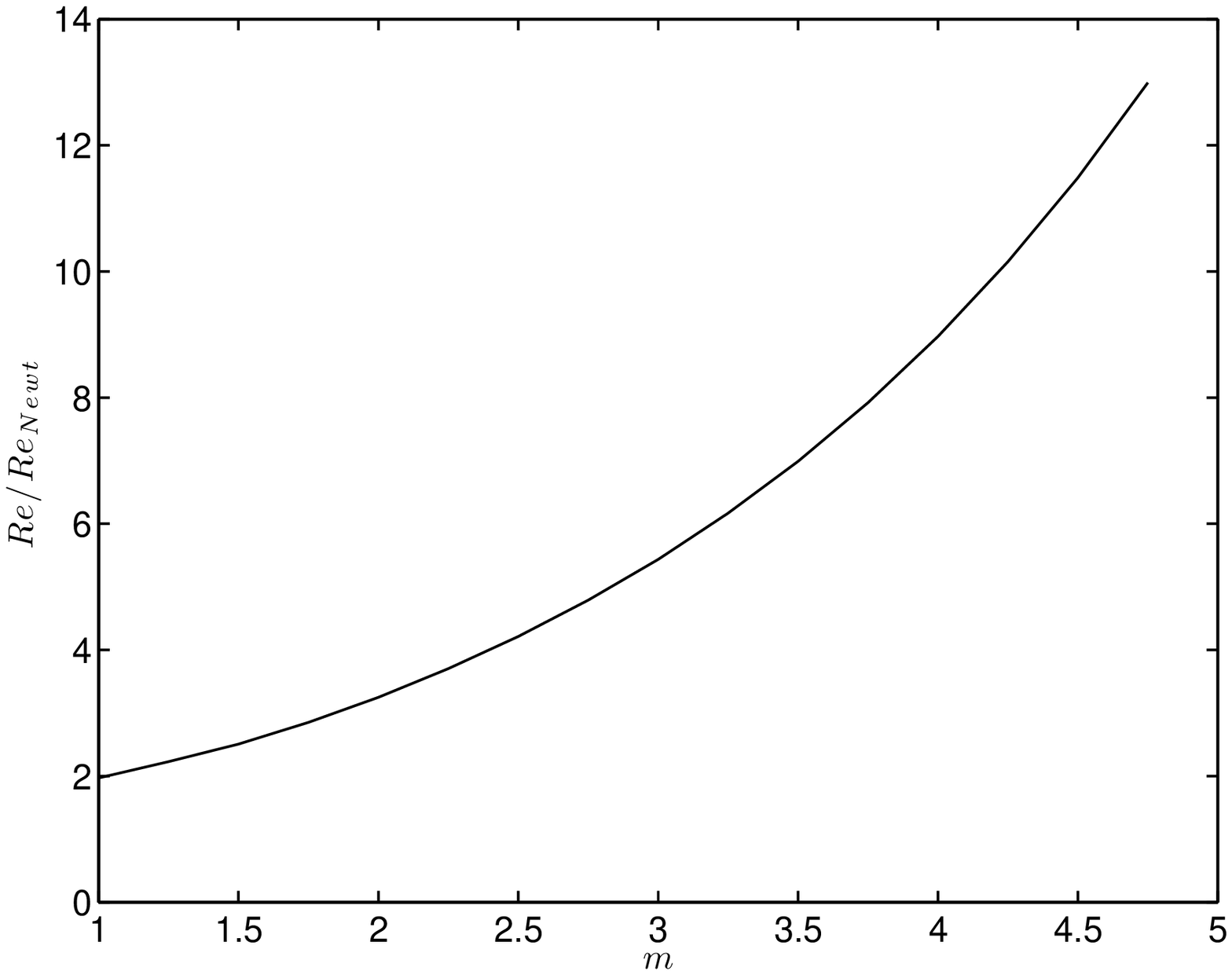}}
(b)\scalebox{0.5}{\includegraphics{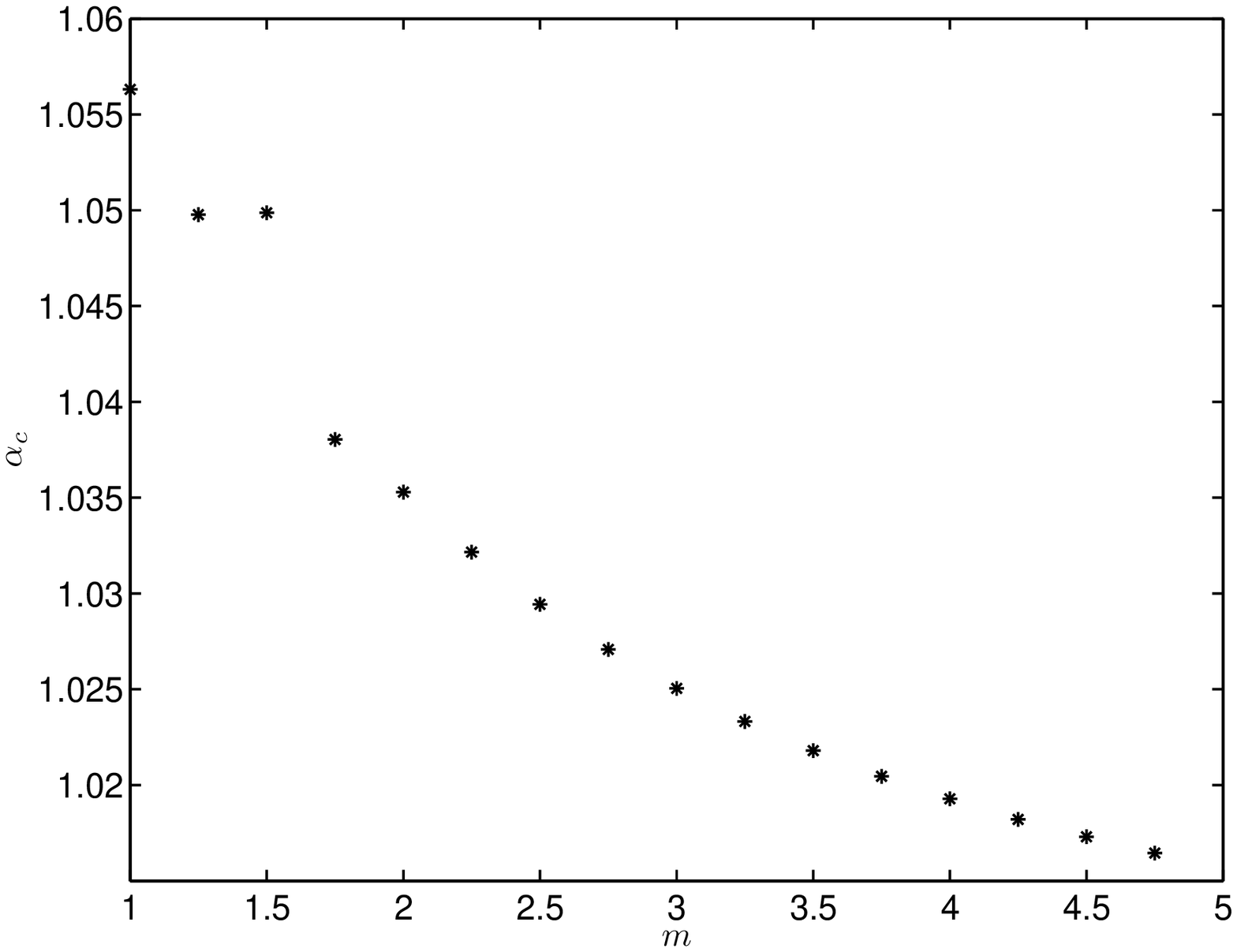}}
\end{center}
\caption{(a) Normalized critical $Re$ for varying power law index $m$ with $Pl=10000$. (b) Critical wave number. All calculations with $\epsilon=0.002$.}
\label{fig:resultsregm}
\end{figure}

Stability results for the regularized Casson model are presented in the following section. First we introduce the ``\emph{plasticity number}'' defined as:

\[Pl=BRe=\frac{\tau_y\rho L^2}{\mu^2}.\]
Note that $Pl$ only depends on the rheological properties of the fluid and the geometry of the problem, thus as we increase $Re$ in our analysis $Pl$ will remained fixed. All the results that follow show the critical Reynolds number for the regularized model normalized with respect to $Re_{Newt}=5772.2$. In Figure \ref{fig:resultsregpl}a we fix $m=2.5$ and with the increase $Pl$ we enhance the stability. These results are not surprising, we are mimicking a  viscoplastic fluid which is known, at least numerically, to be linearly stable \cite{Nouar07b} with a pseudo-plastic fluid with a very large zero-shear rate viscosity. The use of this type of non-linear viscosity has also been shown to enhance stability \cite{Nouar07} which is also represented in Figure \ref{fig:resultsregm}a where we fix $Pl=10000$ and vary the power-law index $m$. Figures \ref{fig:resultsregpl}b and \ref{fig:resultsregm}b show the corresponding critical wave number for increasing plasticity number and increasing power law index respectively. Note that these values remain in close proximity to $\alpha=1.02$ which is the critical wave number for the Newtonian case.

%%%%%%%%%%%%%%%%%%%%%%%%%%%%%%%%%%%%%%%%%%%%%%%%%%%%%%%%%%%%%%%%%%%%%%%%
%%%%%%%%%%%%%%%%					Linear stability analysis full model				%%%%%%%%%%%%%%%%%%%
%%%%%%%%%%%%%%%%%%%%%%%%%%%%%%%%%%%%%%%%%%%%%%%%%%%%%%%%%%%%%%%%%%%%%%%%
\section{Linear stability analysis for Elastoviscoplastic model}

In this section we study the linear stability for the elastoviscoplatic model presented in equations (\ref{eqnmotion})-(\ref{phieqn}).

\subsection{Baseflow profile}

For the base flow we consider unidirectional shear flow of the form $\Ub=(U(y),0,0)$, thus equations (\ref{eqnmotion})-(\ref{phieqn}) reduce to:

\begin{eqnarray}
&&\frac{\dd}{\dd y}\left(\mu_r\frac{\dd U}{\dd y}\right)+\frac{\dd T_{xy}}{\dd y}=Re\frac{\dd p}{\dd x} \label{pmmbaseU}\\
&&T_{xx}=2We\mu(|U'|)\Phi U'T_{xy}\label{pmmbaseTxx}\\
&&T_{xy}=\mu(|U'|)U'\label{pmmbaseTxy}\\
&&T_{yy}=0\label{pmmbaseTyy}\\
&&\Phi=\frac{f(T)}{f(T)+g(T)}\label{pmmbasephi}
\end{eqnarray}
where $T=\sqrt{1/2\left(T_{xx}^2+2T_{xy}^2\right)}$. Substituting (\ref{pmmbaseTxy}) into (\ref{pmmbaseU}) we get:
\begin{equation}
\frac{\dd}{\dd y}\left((\mu_r+\mu(|U'|))\frac{\dd U}{\dd y}\right)=Re\frac{\dd p}{\dd x}\label{pmmbaseflowUfull}
\end{equation}

This equation is solved in exactly the same way as (\ref{eqn:baseflowreg}) in Section \ref{baseflowreg}. Once we have $\gammadot=|U'|$ from (\ref{pmmbaseflowUfull}) we solve the following nonlinear equation for $T_{xx}$:
\begin{equation}
(f(T)+g(T))T_{xx}-2We\left(\mu(|U'|)U'\right)^2f(T)=0\label{Txxnonlineqn}
\end{equation}
where we have used the definition of $T_{xy}$ in (\ref{pmmbaseTxy}).  We now solve equation (\ref{Txxnonlineqn}) using \emph{fzero} in \trademark{MATLAB} for each point $y$ in our discrete domain. Finally we can construct $\Phi$ using (\ref{pmmbasephi}).

\begin{figure}
\begin{center}
\scalebox{0.8}{\includegraphics{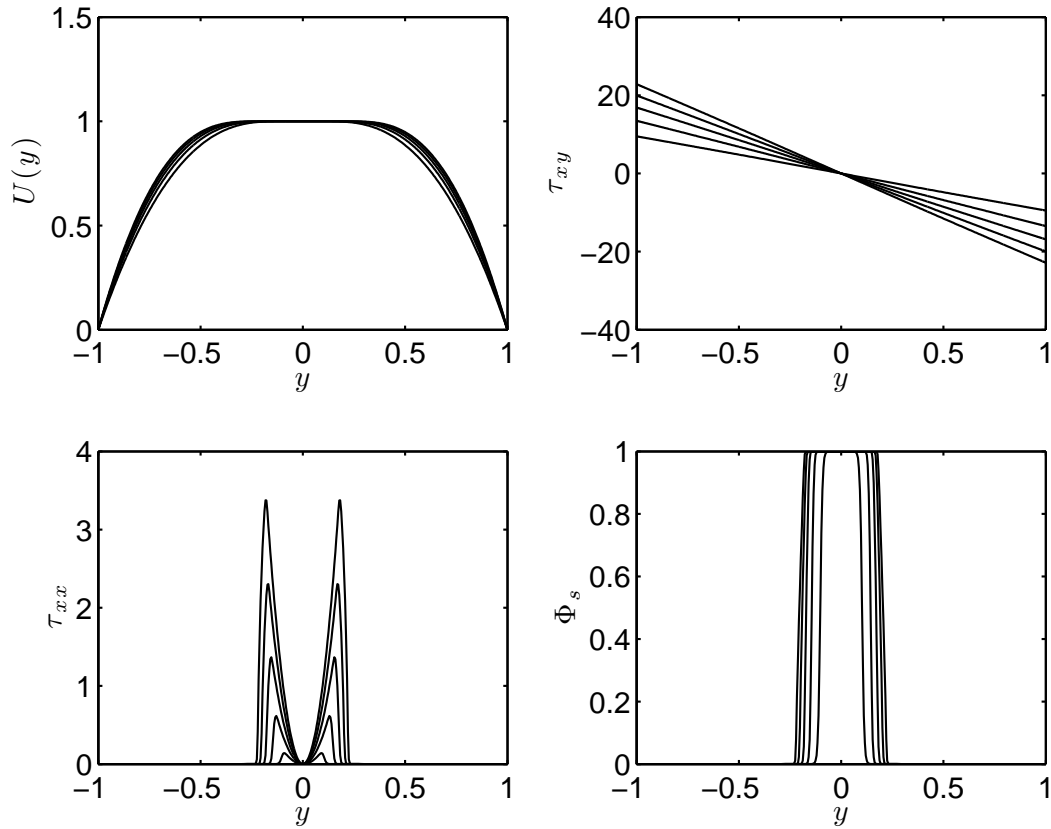}}
\end{center}
\caption{Base flow profile for the elastoviscoplastic model for increasing Bingham number $B$. From top left to bottom right: velocity, $U(y)$; shear stress $\tau_{xy}$; normal stress, $\tau_{xx}$ and steady state structural variable $\Phi_s$. Parameters for this calculation: $\epsilon=0.002$, $We=1$, $m=2.5$, $K_r=K_d=0.3$, $w=0.1$ and $\mu_r=0.1$.}
\label{fig:baseflowpmm}
\end{figure}

In Figure \ref{fig:baseflowpmm} we present the solution of equations (\ref{pmmbaseU})-(\ref{pmmbasephi}) for increasing Bingham numbers and fixed Weissenberg number $We=1$. As we can see the velocity profile presents the pseudo plastic behaviour which in this case, in addition of a intensely stratified viscosity, we have a viscoelastic plug, due to the presence of non-zero normal stresses inside this region. We should note the existence of a thin region ($O(w)$) where both solid (viscoelastic) and fluid phases coexist as it's clearly seen in the plot of $\Phi_s$.

\subsection{Orr-Sommerfeld equation for Elastoviscoplastic model}

As before, we consider an infinitesimal perturbation $(\varepsilon\ub', \varepsilon p', \varepsilon \taub, \varepsilon \Phi')$ superimposed onto the base flow and the field equations are linearized around $(\Ub, P, \Tb, \Phi)$, thus we have:

\begin{eqnarray}
&&\frac{\partial \ub'}{\partial t}+(\ub'\cdot{\bm \nabla})\Ub+(\Ub\cdot{\bm \nabla})\ub'=-\nabla p'+\frac{\mu_r}{Re} {\bm \nabla} \cdot \gammabdot(\ub')+\frac{1}{Re} {\bm \nabla} \cdot\taub' \label{eqnmotionpertpmm}\\
&&{\bm \nabla} \cdot\ub'= 0 \label{conteqnpertpmm},\\
&&\taub'+We(\Phi\mu(\stackrel{\triangledown_{\ub'}}{\Tb}+\stackrel{\triangledown_{\Ub}}{\taub'})+(\Phi\tilde{\mu}+\Phi'\mu)\stackrel{\triangledown_{\Ub}}{\Tb}) = \mu_t \gammabdot(\ub'), \label{constpertpmm}
\end{eqnarray}

\begin{eqnarray}
\frac{\partial \Phi}{\partial t}+(\ub'\cdot{\bm \nabla})\Phi+(\Ub\cdot{\bm \nabla})\Phi'&=&-\left(\left(\frac{\partial g(T)}{\partial T_{xx}}+\frac{\partial f(T)}{\partial T_{xx}}\right)\tau'_{xx}+\left(\frac{\partial g(T)}{\partial T_{xy}}+\frac{\partial f(T)}{\partial T_{xy}}\right)\tau'_{xy}\right)\Phi\nonumber \\
&+&\left(\frac{\partial f(T)}{\partial T_{xx}}\tau'_{xx}+\frac{\partial f(T)}{\partial T_{xy}}\tau'_{xy}\right)-(g(T)+f(T))\Phi'.\label{phieqnpertpmm}
\end{eqnarray}
As there is no version of Squire Theorem for this type of models for simplicity we restrict ourselves to two-dimensional perturbations below. We introduce a stream function $\Psi$, such that: $u' = \Psi_y,~v'=-\Psi_x$. We consider modal linear disturbances of the form:
\begin{equation}
\Psi=\phi(y)\ee^{\ii(\alpha x-\omega t)},~~ \tau'_{ij}=\tau_{ij}(y)\ee^{\ii(\alpha x-\omega t)},~~ij = xx,~xy,~yx,~yy,~~\Phi'=\varphi(y)\ee^{\ii(\alpha x-\omega t)},.
\end{equation}
Denoting $D = \frac{\dd}{\dd y}$, the eigenvalue problem for $\omega$ is:
\begin{eqnarray}
\ii\omega (\alpha^2-D^2)\phi
&=&\frac{\mu_r}{Re}(\alpha^2 - D^2)^2 \phi + \ii\alpha [D^2U + U (\alpha^2-D^2)] \phi  \nonumber\\
& &+\frac{1}{Re}[\ii\alpha D(\tau_{xx} - \tau_{yy}) + (D^2 + \alpha^2) \tau_{xy} ]\label{phieqnpmm} \\
\nonumber \\
\ii\omega We\Phi\mu \tau_{xx}
&=&
\left[1+\ii\alpha We\Phi\mu U \right] \tau_{xx} -2We\Phi\mu DU \tau_{xy} 
-\ii\alpha We\Phi\mu DT_{xx} \phi -2We\Phi\mu T_{xy}D^2 \phi \nonumber \\
& &
-2\ii\alpha \left[We\Phi\mu T_{xx}+\mu\right]D\phi - 2\Phi\mu DU T_{xy}\varphi \label{tau1eqn}\\[0.5ex]
\ii\omega We\Phi\mu \tau_{xy}
&=&
\left[1+\ii\alpha We\Phi\mu U\right]\tau_{xy} - We\Phi\mu DU \tau_{yy} - \mu_tD^2\phi \nonumber \\
& &-\left[ \ii\alpha We\Phi\mu DT_{xy} + We\Phi\mu\alpha^2 T_{xx} + \alpha^2\mu_t\right] \phi \label{tau2eqn}\\[0.5ex]
\ii\omega We\Phi\mu \tau_{yy} &= & \left[1+\ii\alpha We\Phi\mu U\right]\tau_{yy} + 2\ii\alpha\mu D\phi-2We\Phi\mu\alpha^2T_{xy}\phi . \label{tau3eqn}\\
\ii\omega \varphi 
&=&
\ii\alpha D\Phi\phi+\left[\ii\alpha U+g(T)+f(T)\right]\varphi + \left[(K_r-K_d)\Phi-K_r \right]F(y)\mbox{sign}(DU)\tau_{xy} \nonumber \\
& &
+ \left[(K_r-K_d)\Phi^2-K_r\Phi \right]F(y)We\mu |DU|\tau_{xx}, \label{varphieqn}
\end{eqnarray}
where
\begin{equation}
F(y)=\frac{1-\tanh^2\left(\frac{T-B}{w}\right)}{w\sqrt{2(We\mu DU\Phi)^2+1}}
\end{equation}
and with boundary conditions:
\begin{equation}
\phi = D\phi = 0,~\mbox{ at } y=\pm1. \label{bcspmm}
\end{equation}

As it is customarily done in linear stability analysis for viscoelastic flows we introduce the elasticity number defined as:
\[\elast=\frac{We}{Re},\]
which is the ratio of kinematic viscosity and relaxational diffusivity: $\elast$ depends only on the properties of the fluid and the flow geometry. For the rest of the paper we replace $We$ by $\elast Re$.
%------------------------------------------------------------------------------------
\subsubsection{One-dimensional stability}

In this section we consider one-dimensional disturbances of (\ref{phieqnpmm})-(\ref{varphieqn}). In order for $\phi$ to satisfy boundary conditions (\ref{bcspmm}) we proceed in the same way as for the regularized viscoplastic fluid, by letting the perturbation to the flow rate $\int_{-1}^1u(y)\dd y$ to be zero and we impose a uniform pressure gradient in the $x$-direction. Letting $\alpha=0$ in (\ref{phieqnpmm})-(\ref{varphieqn}) we are left with the following eigenvalue problem:

\begin{eqnarray}
D^2\tau_{xy}&=&-\ii Re \omega D^2\phi -\mu_rD^4\phi, \label{pmmphieqnalp0}\\
\left(1-\ii \elast Re \Phi \mu \omega \right)\tau_{xx}&=& 2\elast Re\Phi\mu DU \tau_{xy} 
+2\elast Re\Phi\mu T_{xy}D^2 \phi+2\Phi\mu DU T_{xy}\varphi , \label{pmmtauxxeqnalp0}\\
\left(1-\ii\elast Re \Phi \mu \omega \right)\tau_{xy}&=&\mu_tD^2\phi. \label{pmmtauxyeqnalp0}\\
\left(1-\ii\elast Re \Phi \mu \omega \right)\tau_{yy}&=&0, \label{pmmtauyyeqnalp0}\\
\left(\ii\omega-g(T)-f(T) \right) \varphi&=& \left[(K_r-K_d)\Phi-K_r \right]F(y)\mbox{sign}(DU)\tau_{xy} \nonumber \\
& &
+ \left[(K_r-K_d)\Phi^2-K_r\Phi \right]F(y)\elast Re\mu |DU|\tau_{xx}\label{pmmphiseqnalp0}
\end{eqnarray} 

We should point out that from equations (\ref{pmmtauxxeqnalp0})-(\ref{pmmphiseqnalp0}) a continuous spectrum exists and this consists of purely imaginary eigenvalues in  the strip $[-1/\elast Re \max(\mu),-\infty)$ (note that for our choice of regularization parameter $\epsilon$, we always have $(K_d+K_r)>1/\elast Re \max(\mu)$).  One has to be careful when considering this part of the spectrum, note that as $\Phi\rightarrow 0$ then $\omega_i\rightarrow -\infty$ , clearly these eigenvalues are always damped, thus we consider only the case $\tau_{yy}=0$. Also note that (\ref{pmmtauxxeqnalp0}) and(\ref{pmmphiseqnalp0}) decouple from (\ref{pmmphieqnalp0}) and (\ref{pmmtauxyeqnalp0}), then we multiply (\ref{pmmphieqnalp0}) by the conjugate of $\phi$, say $\phi^*$,  (\ref{pmmtauxyeqnalp0}) by $\tau_{xy}^*$ and integrate by parts to get:

\begin{eqnarray}
&&\int_{-1}^1\tau_{xy}D^2\phi^*\dd y=\ii Re \omega \int_{-1}^1|D\phi|^2\dd y -\mu_r\int_{-1}^1|D^2\phi|^2\dd y, \label{alp0phieqn}\\
&&\int_{-1}^1\left(1-\elast Re \Phi \mu \omega \right)|\tau_{xy}|^2\dd y=\int_{-1}^1\mu_tD^2\phi\tau_{xy}^*\dd y. \label{alp0txyeqn}
\end{eqnarray} 

Expanding in real and imaginary parts and applying the mean value theorem for integrals as necessary we have the following equation for the imaginary part of $\omega$, 

\begin{equation}
\omega_I=-\frac{\int_{-1}^1|\tau_{xy}|^2\dd y+\mu_{t_1}\mu_r\int_{-1}^1|D^2\phi|^2\dd y}{\mu_{t_1}Re\int_{-1}^1|D\phi|^2\dd y+\elast Re \bar{\Phi}\bar{\mu}\int_{-1}^1|\tau_{xy}|^2\dd y}<0\label{onedstabeqn}
\end{equation}
where
\[\mu_{t_1}=\mu_t(\xi_1),~~\bar{\Phi}=\Phi(\xi_2),~~\mbox{and}~~\bar{\mu}=\mu(\xi_3)~~~\mbox{for some}~~\xi_1,\xi_2,\xi_3\in[-1,1].\]

Therefore, (\ref{phieqnpmm})-(\ref{varphieqn}) is linearly stable to one-dimensional perturbations.
%----------------------------------------------------------------------------------------------------------------------------------------
\subsubsection{Code validation}

\begin{figure}
\begin{center}
\scalebox{0.5}{\includegraphics{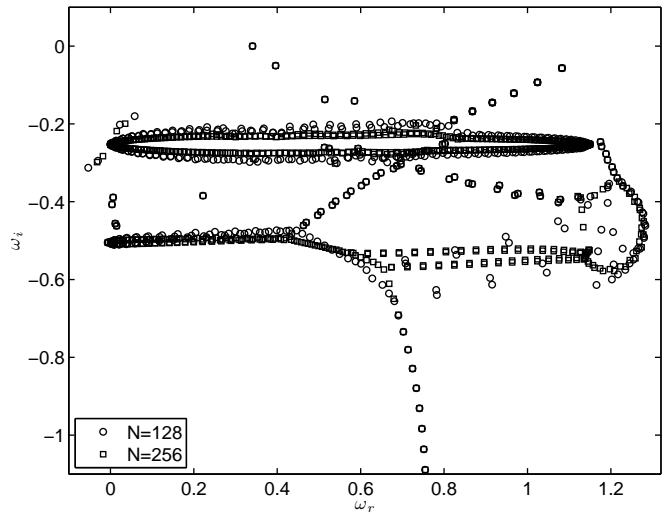}}
\end{center}
\caption{Code validation. Spectrum for Oldroyd-B fluid, $N=128$ (o), $N=256$ ($\square$). For $Re=3960$, $\elast=10^{-3}$ ($We=3.96)$, $\beta=0.5$ and $\alpha=1.15$.}
\label{fig:validationOB}
\end{figure}

As a first test for our code we consider the case $\mu_r=0.5$, $\Phi=\mu=\mu_t\equiv1$ and $\varphi\equiv0$ in (\ref{phieqnpmm})-(\ref{varphieqn}), by doing this we recover the Oldroyd-B model and compare our results with the work by Sureshkumar and Beris in \cite{Suresh95}. We set $Re=3960$, $We=3.96$ (or $\elast=10^{-3}$), $\alpha=1.15$ and have used $N=128$ and $N=256$ to solve for the spectrum. We show the results in Figure \ref{fig:validationOB},  the least stable eigenvalue is $\omega=0.3409+ 1.9888\times 10^{-7}\ii$, the value reported in \cite{Suresh95} is $\omega=0.3409+ 1.9696\times 10^{-7}\ii$.  We have convergence of the discrete spectrum and the difference of the least stable eigenvalue between computations is $2.7\times 10^{-12}$.

We now turn our attention to the full model (\ref{phieqnpmm})-(\ref{varphieqn}). For the rest of the section we consider the wave speed $c=\omega/\alpha$ as our eigenvalue. The first thing to note is the location of the continuous spectrum, from (\ref{tau1eqn})-(\ref{varphieqn}). For simplicity let us define the Deborah function as $De(y)=\elast\Phi\mu$. We can see that the real part of the continuous spectrum is in the line $c_r= U$ and the imaginary is in the line $c_i\in [a,\infty)$, where:

\[a=\max\{2K_d,2K_r,WeB/\epsilon\}.\]

\begin{figure}
\begin{center}
(a)\scalebox{0.5}{\includegraphics{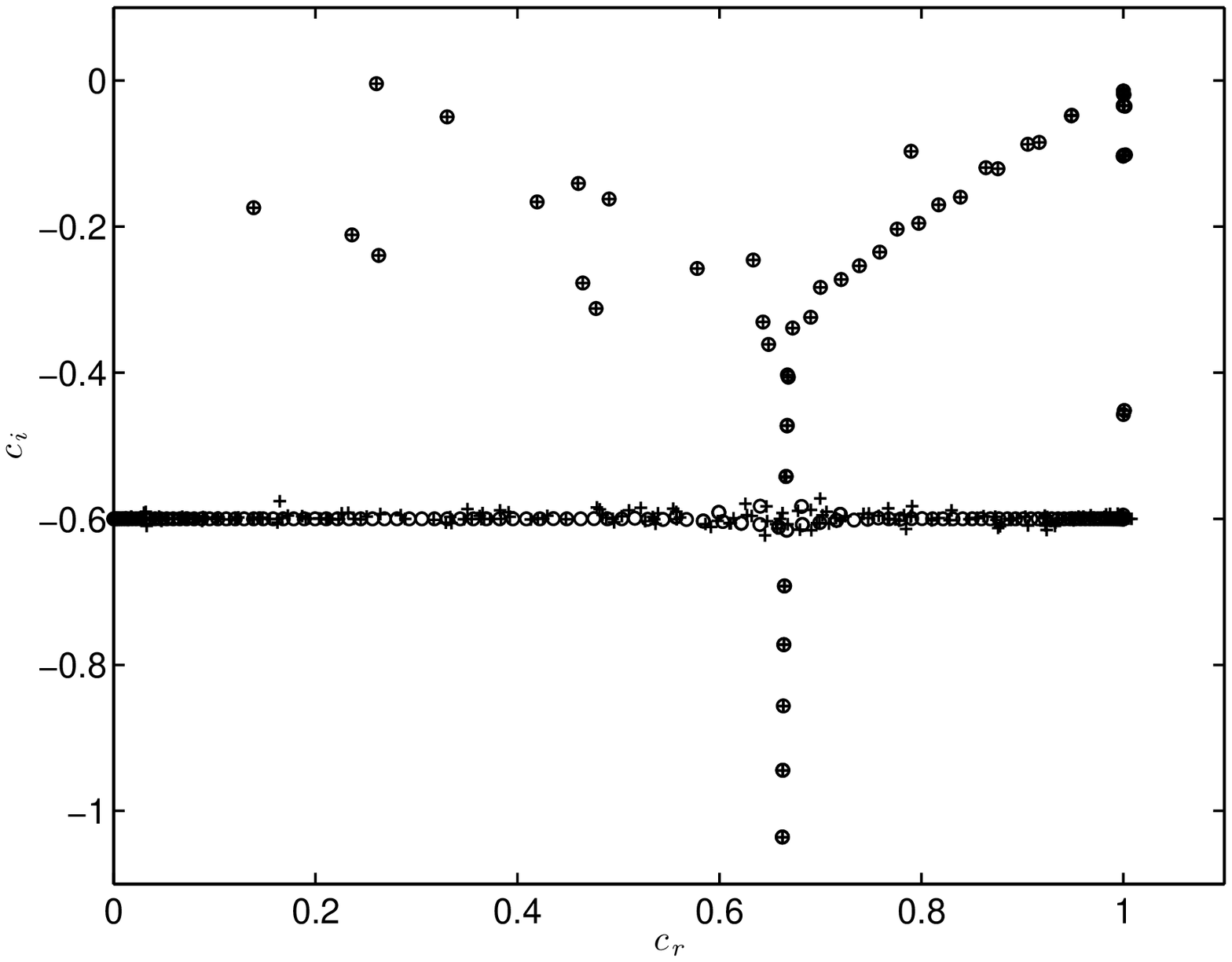}}\\
(b)\scalebox{0.5}{\includegraphics{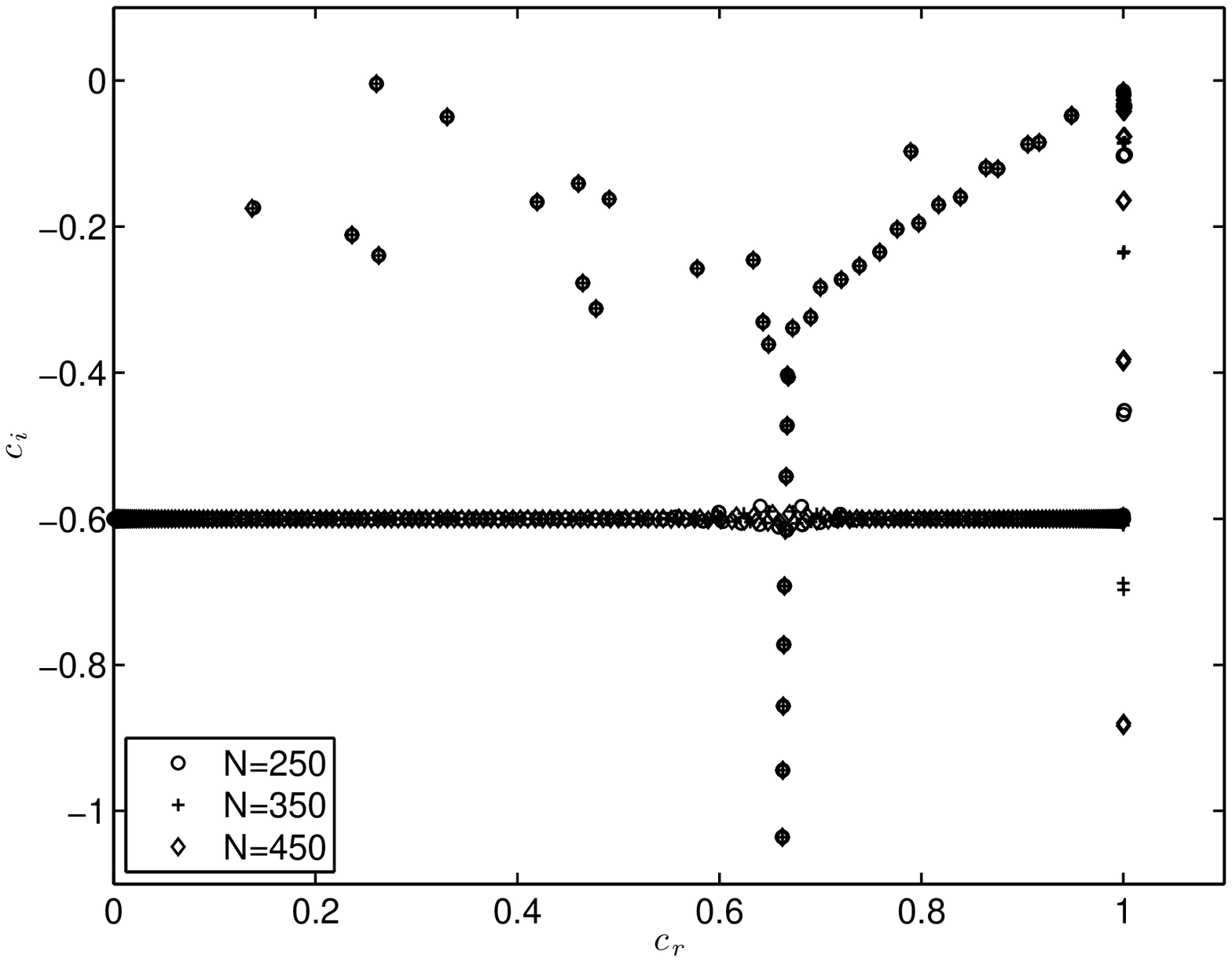}}
\end{center}
\caption{Code validation for elastoviscoplatic model. a) Spectrum for full model (o) and for regularized model using $\Phi_r$ (+) with $N=250$. b) Spectrum for full model with $N=250$ (o), $N=350$ ($+$), and $N=450$($\Diamond$). For $Re=10000$ and $\alpha=1$. Parameters for these calculations: $\epsilon=0.002$, $\elast=10^{-4}$, $m=2.5$, $K_r=K_d=0.3$, $w=0.1$ and $\mu_r=0.1$.}
\label{fig:validationpmm}
\end{figure}

Clearly, the continuous spectrum is always stable but one has to be careful when doing computations. Note that for $y$ in the pseudo-plug region (where $T<B$), $De(y)$ is very large (recall that $\mu\sim B/\epsilon$), once we begin to approach the yield limit $De(y)$ decays to zero very fast  as $\Phi\rightarrow0$ and $\mu\rightarrow1$. This means that the continuous spectrum will approach zero when $y$ is on the pseudo-plug region and it will be pulled down very fast towards minus infinity as $T$ approaches $B$. This could cause severe numerical problems, $\Phi$ is actually zero when the fluid is fully yielded. We discretize the problem using the same technique as in Section \ref{subsec:validationmodOrr-Som}.  In order to prove the convergence of our code we compare the spectrum of a regularized version of (\ref{phieqnpmm})-(\ref{varphieqn}) with the one of the full model using two different eigenvalue solvers in \trademark{MATLAB}. Let
\begin{equation}
\Phi_r (y)= \left\{
\begin{array}{lcl}
\Phi(y) & \mathrm{for} & \Phi(y)>\vartheta\\
\vartheta & \mathrm{for} & \Phi(y)\leq \vartheta,\\
\end{array}
\right.
\end{equation}
where $\vartheta\ll1$ is a small parameter. We replace $\Phi$ with $\Phi_r$ in $De(y)$ and solve the generalized eigenvalue problem using the function \emph{eig} in \trademark{MATLAB}, we choose $\vartheta$ such that we get finite results. We then solve the full problem (\ref{phieqnpmm})-(\ref{varphieqn}) using \emph{eigs} in \trademark{MATLAB}, this function provides the reverse communication required by the Fortran library ARPACK which is based upon an algorithmic variant of the Arnoldi process called the Implicitly Restarted Arnoldi Method. This function can calculate $k$ eigenvalues based on a user defined $\sigma$, thus we choose $k$ such that we get finite results and $\sigma$ is taken to be the least stable eigenvalue of the regularized Casson model calculated in Section  \ref{subsec:validationmodOrr-Som}. We present the results of these calculations with $N=250$ in Figure \ref{fig:validationpmm}a, the spectra of both problems overlap each other. Note the presence of the almost undisturbed spectrum of the generalized Newtonian problem ($A$-branch, $P$-branch, $S$-branch and $R$-branch), the continuous spectrum corresponding to equation (\ref{varphieqn}) centered at $c_i=(K_d+K_r)/\alpha^2$ (we have assumed $K_d=K_r$ for this example). Because the choice of parameters (see caption in figure), the pseudo-plug region is very small we have only a few eigenvalues corresponding to $c_r=U\sim 1$ and $c_i=-1/\alpha^2De(y)Re$.  Note that there exists a set of discrete eigenvalues which emanates from the region close to $-1/\alpha^2De(0)Re$ which distorts the $A,~P,~R$ branches of the modified Orr-Sommerfeld operator. The difference on the least stable eigenvalue between the problem using $\Phi_r$ and the one using $\Phi$ is $5.1101\times10^{-6}$. This result is not surprising due to the fact that this eigenvalue belongs to the wall modes ($A$-branch), there the fluid is fully yielded and it does not feel the effect of elasticity, when we use $\Phi_r$ close to the wall we can assume that $\elast\sim O(10^{-8})$ for this choice of parameters. In view of these results we use the function \emph{eigs} for the full discretized problem (\ref{phieqnpmm})-(\ref{varphieqn}). In Figure \ref{fig:validationpmm}b we present convergence results for $N=250,~350,~450$, the change on the least stable eigenvalue is on the sixth significant figure. For the rest of the paper we fix $N=450$ unless otherwise stated.

%\begin{figure}
%\begin{center}
%\scalebox{0.5}{\includegraphics{figuredeborahfunc.eps}}
%\end{center}
%\caption{Spectrum for regularized shear-thinning Casson's model, $N=150$ (o), $N=200$ ($+$), and $N=250$(diamonds). For $Re=10000$ and $\alpha=1$.}
%\label{fig:deborahRe}
%\end{figure}

%%%%%%%%%%%%%%%%%%%%%%%%%%%%%%%%%%%%%%%%%%%%%%%%%%%%%%%%%%%%%%
%%%%%%%%%%%			Numerical results							%%%%%%%%%%%%%%%%%%%
%%%%%%%%%%%%%%%%%%%%%%%%%%%%%%%%%%%%%%%%%%%%%%%%%%%%%%%%%%%%%%

\subsection{Numerical results}

\begin{figure}
\begin{center}
(a)\scalebox{0.5}{\includegraphics{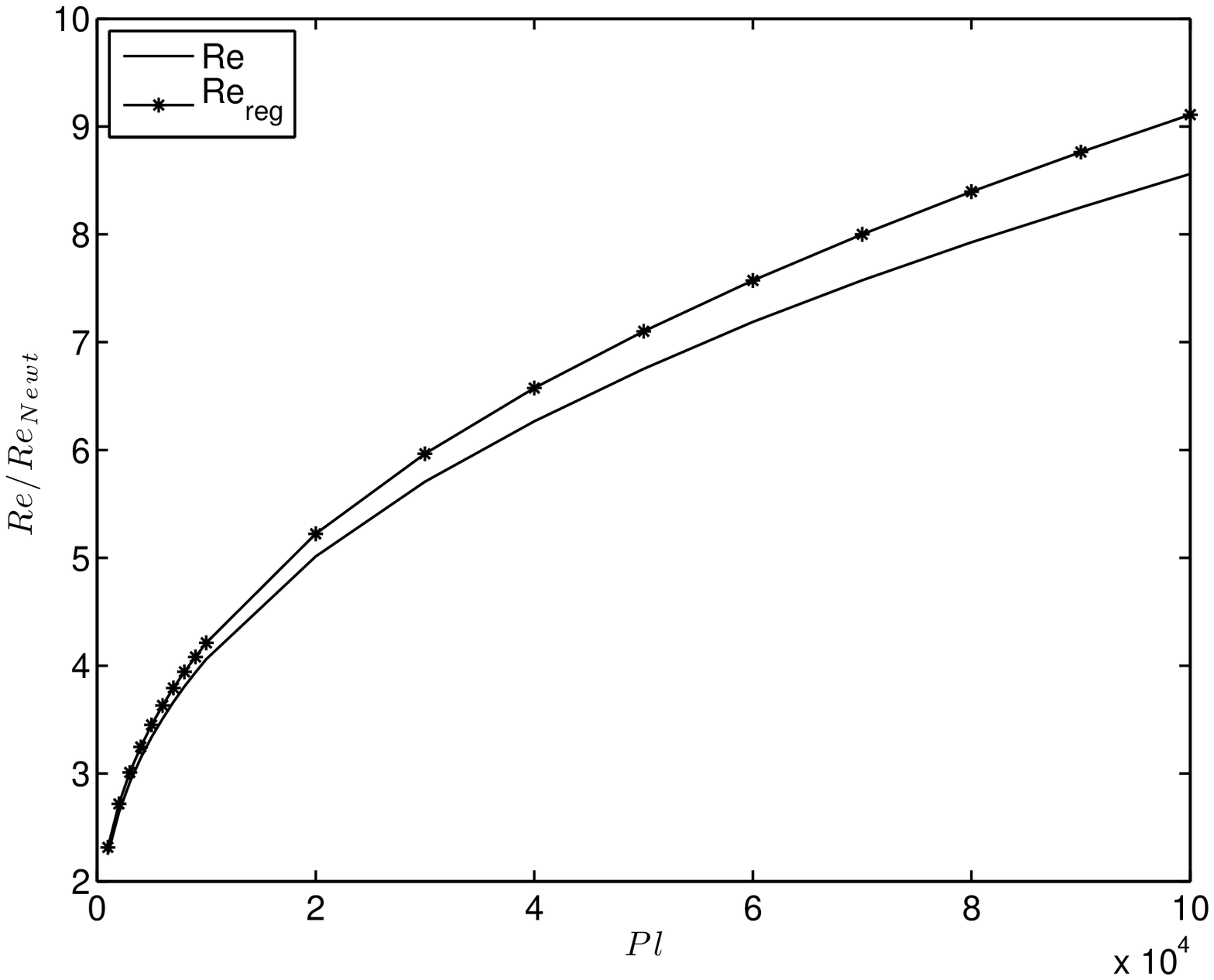}}
(b)\scalebox{0.5}{\includegraphics{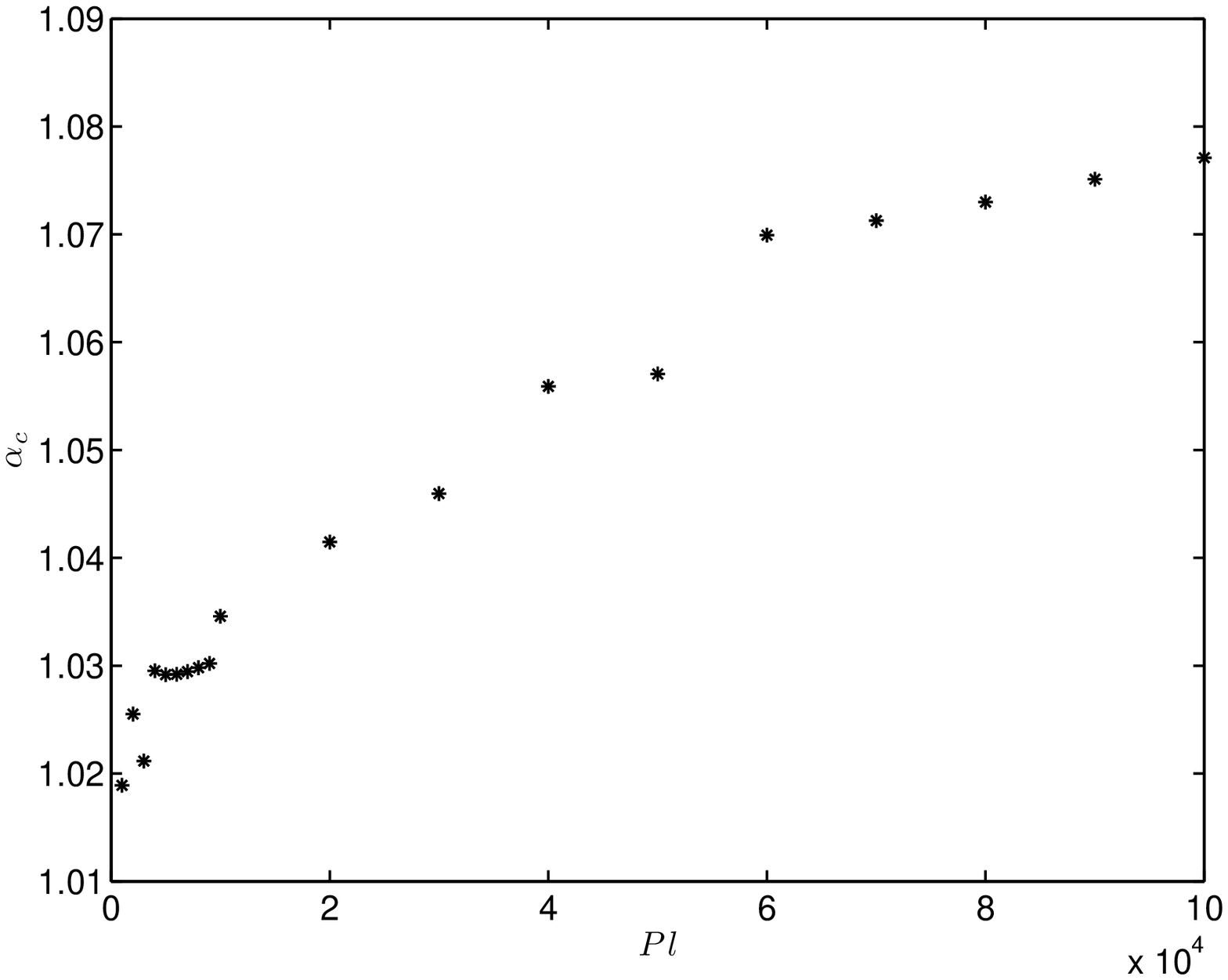}}
\end{center}
\caption{(a) Normalized  critical $Re$ for increasing $Pl$. (b) Critical wave number. Parameters for these calculations: $\epsilon=0.002$, $\elast=10^{-4}$, $m=2.5$, $K_r=K_d=0.3$, $w=0.1$ and $\mu_r=0.1$.}
\label{fig:resultspmmpl}
\end{figure}

Next we present the results of our cacluations. Figure \ref{fig:resultspmmpl}a shows the critical Reynolds number for the elastoviscoplastic model as a function of plasticity number, $Pl$. As with the regularized Casson model the existence of a pseudo-plug region (stratified viscosity) is suffciente to greatly enhance the stability of the flow. The critical Reynolds number appears to be a monotone increasing function of plasticity number, just as with the regularized viscoplastic model. Clearly, in Figure \ref{fig:resultspmmpl}b we can see that the behaviour of the critical wave number $\alpha_c$ is very similar to the wave number of the Casson model. We should also note that the inclusion of a highly viscous viscoelastic fluid as a plug destabilises the flow when comparing it with the regularized model (shown as the dotted curve in Figure \ref{fig:resultspmmpl}a). In relative terms, when $Pl=1000$ the critical Reynolds number for the elastoviscoplastic model is 2.66\% smaller than the critical Reynolds number for regularized model. For $Pl=10^5$ this percentage increases to around 6\%, this is due to the fact that the pseudo-plug and solid-fluid regions increase and are closer to the wall. 

\begin{figure}
\begin{center}
(a)\scalebox{0.5}{\includegraphics{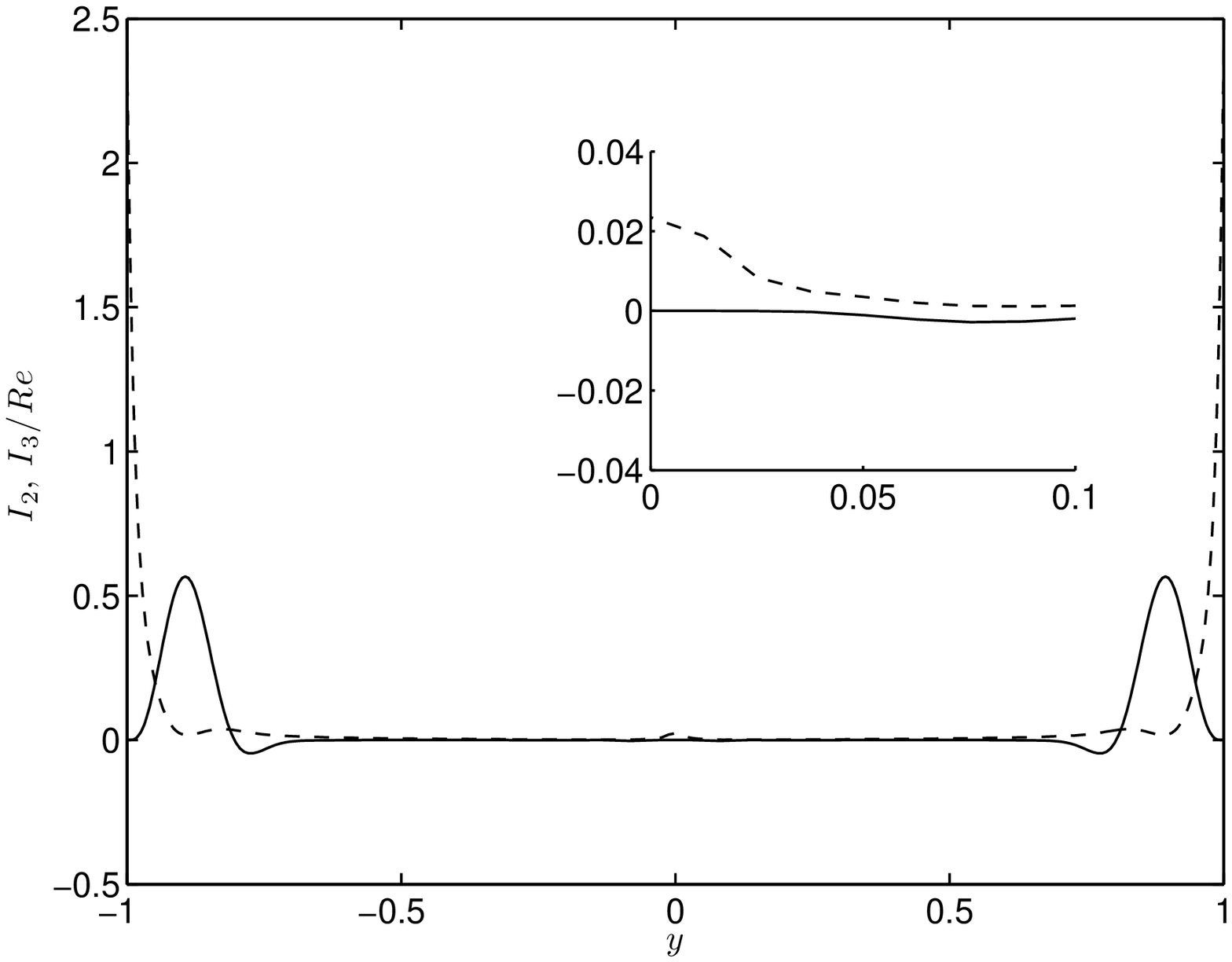}}
(b)\scalebox{0.5}{\includegraphics{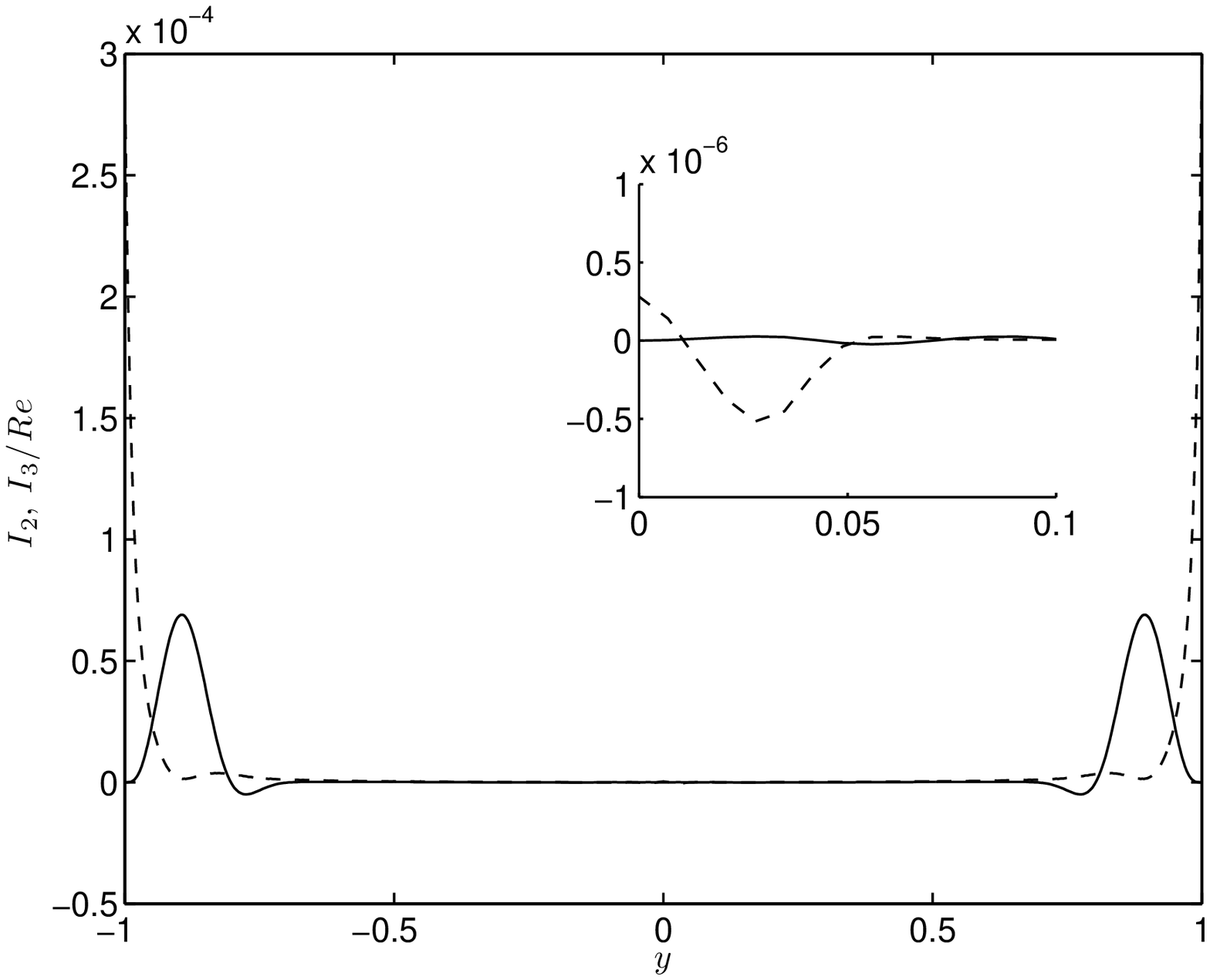}}
\end{center}
\caption{Energy budget for $Re=13364.41$ and $Pl=1000$. (a)Regularized Casson model, $\Gamma^+=\Gamma^-=7.152\times10^{-3}$ (b) Elastoviscoplastic model, $\Gamma^+=7.351\times10^{-3},~\Gamma^-=7.033\times10^{-3}$. Parameters for these calculations: $\epsilon=0.002$, $\elast=10^{-4}$, $m=2.5$, $K_r=K_d=0.3$, $w=0.1$ and $\mu_r=0.1$.}
\label{fig:energybudget}
\end{figure}

As a way to understand better the effects of the elastoviscoplasticity on the instability mechanism we will turn our attention to the analysis of the distribution of the production and dissipation of disturbance kinetic energy across the channel. The linearized Reynolds-Orr equation is derived by multiplying equation (\ref{phieqnpmm}) (equation (\ref{Orr-Somreg}) for the regularized model) by the conjugate of the stream function perturbation (or velocity perturbation in (\ref{Orr-Somreg})) and integrating over a periodic domain. This can be written as:
\begin{equation}
\frac{d\langle I_{1,j}\rangle}{dt}=\langle I_{2,j}\rangle-\frac{1}{Re}\langle I_{3,j}\rangle \label{Reynolds-Orr}
\end{equation}
where $j=$regularized (reg) or elastoviscoplastic (ev) and $\langle \cdot \rangle=\int_{-1}^1\cdot\dd y$ . The left hand side of (\ref{Reynolds-Orr}) is the temporal variation of the averaged kinetic energy, the first term in the right hand side is the exchange of energy between the base flow and the perturbation and the last term in the equation is the rate of energy dissipation. Explicitly we have for the regularized model
\begin{eqnarray}
I_{1,reg}&=&|Dv|^2+\alpha^2|v|^2 \label{I1reg}\\
I_{2,reg}&=&\alpha[DU_{reg}(v_rDv_i-v_iDv_r)] \label{I2reg}\\
I_{3,reg}&=&4\alpha^2\mu|Dv|^2+\mu_t|D^2v+\alpha^2v|^2 \label{I3reg}
\end{eqnarray}
and for the elastoviscoplastic model
\begin{eqnarray}
I_{1,ev}&=&|Dv|^2+\alpha^2|v|^2 \label{I1pmm}\\
I_{2,ev}&=&\alpha[DU_{ev}(\phi_rD\phi_i-\phi_iD\phi_r)] \label{I2pmm}\\
I_{3,ev}&=&\mu_r|D^2\phi+\alpha^2\phi|^2\nonumber\\
&+&\alpha[(\tau_{yy}-\tau_{xx})_rD\phi_i-(\tau_{yy}-\tau_{xx})_iD\phi_r] \nonumber\\
&+&\tau_{xy_r}(D^2\phi_r+\alpha^2\phi_r)+\tau_{xy_i}(D^2\phi_i+\alpha^2\phi_i).\label{I3pmm}
\end{eqnarray}
Following Govindarajan et al. \cite{Govindarajan01}, we compare the normalized space-averaged energy production,
\begin{equation}
\Gamma^+_j=\frac{\langle I_{2,j}\rangle}{\langle I_{1,j}\rangle}\label{energyprod}
\end{equation}
and normalized space-averaged energy dissipation
\begin{equation}
\Gamma^-_j=\frac{\langle I_{3,j}\rangle}{Re\langle I_{1,j}\rangle}.\label{energydiss}
\end{equation}
At criticality these two quantities balance. Positive values of $I_{2,j}$ indicate where in the flow domain energy is supplied from the base flow to the perturbed flow. It should be mentioned that even though that the energy production and energy dissipation balance at critical conditions, the mechanism of instability is governed by $I_{2,i}$ and not by $I_{3,j}$. We refer the reader to \cite{Govindarajan01} for an extensive discussion about Reynolds stress distribution analysis. In Figure \ref{fig:energybudget}a we can see how $\Gamma^+_{reg}$ and $\Gamma^-_{reg}$ balance at critical Reynolds number $Re_{reg}=13364.41$ for $Pl=1000$. The inclusion of an elastic plug an its effect on the stability of the flow is clearly seen in Figure \ref{fig:energybudget}b. We keep the same Reynolds number as for the regularized case. Note that both $\Gamma^+_{ev}$ and $\Gamma^-_{ev}$ are in the same orders of magnitude as before, but the balance is slightly broken. The energy production has increased but note that the energy dissipation has gone negative close to the centre line of the channel (insert figure) whereas for the regularized model is always positive due to the presence of a highly viscous pseudo-plug. This is due to the presence of elastic forces. Even though the elastic plug is confined to a very small region away from the wall, it is enough to break the balance between energy production and dissipation and stability is lost. Experimental evidence of the existence of an elastic plug can be found in \cite{Park10}.

\begin{figure}
\begin{center}
(a)\scalebox{0.5}{\includegraphics{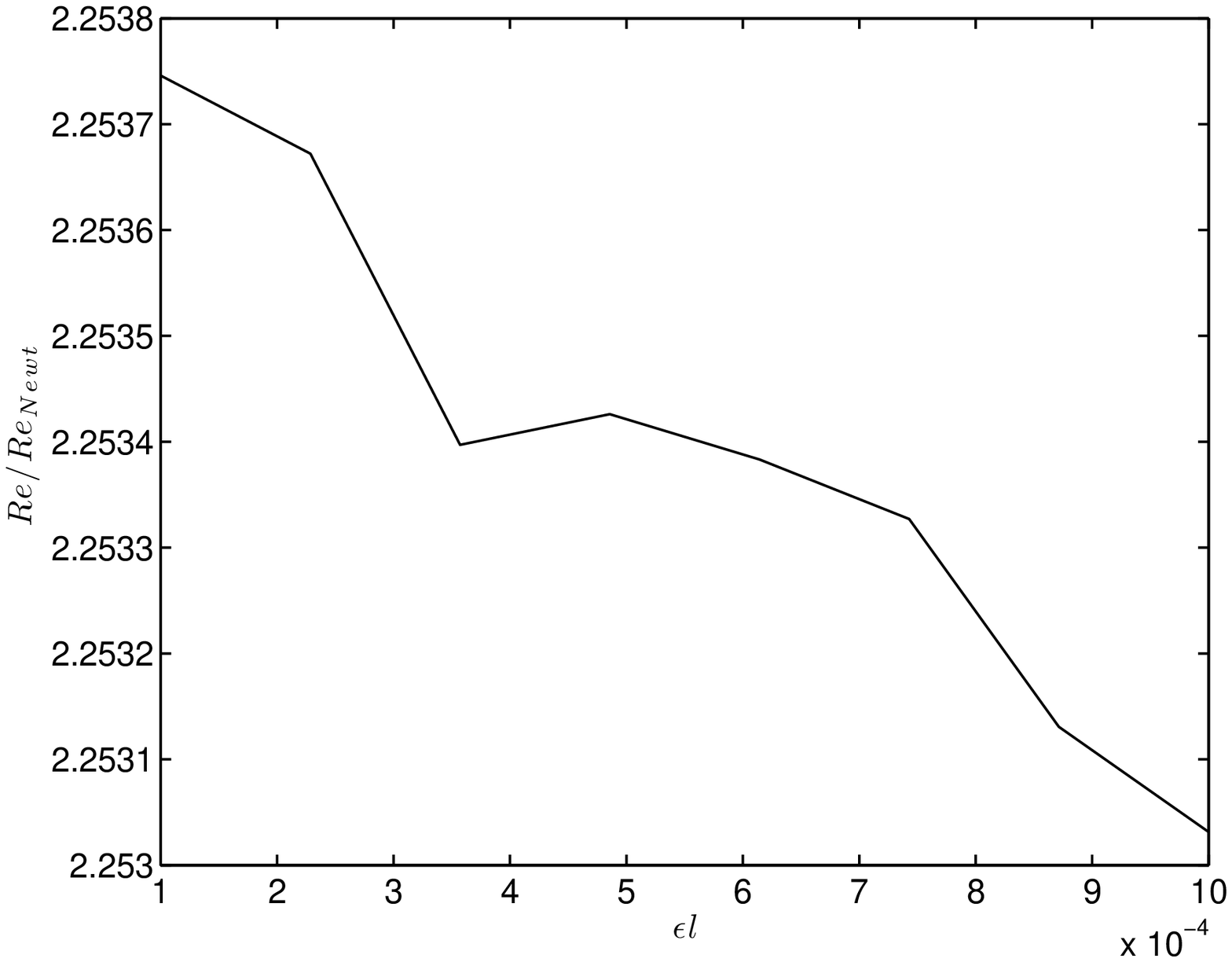}}
(b)\scalebox{0.5}{\includegraphics{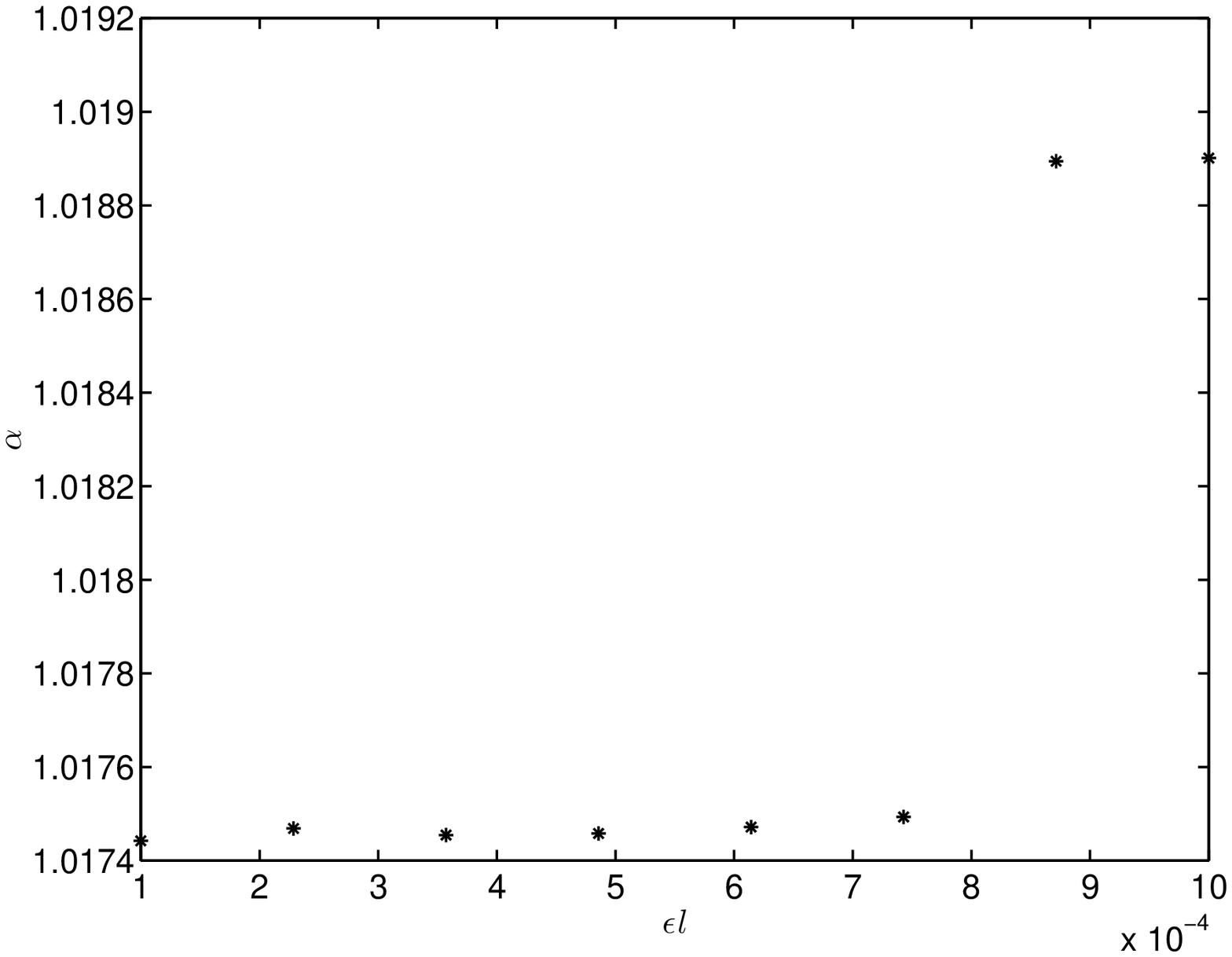}}
\end{center}
\caption{(a) Normalized  critical $Re$ with $Pl=1000$ and increasing $\elast$. (b) Critical wave number. Parameters for these calculations: $\epsilon=0.002$, $m=2.5$, $K_r=K_d=0.3$, $w=0.1$ and $\mu_r=0.1$.}
\label{fig:resultspmmeps}
\end{figure}

\begin{figure}
\begin{center}
(a)\scalebox{0.5}{\includegraphics{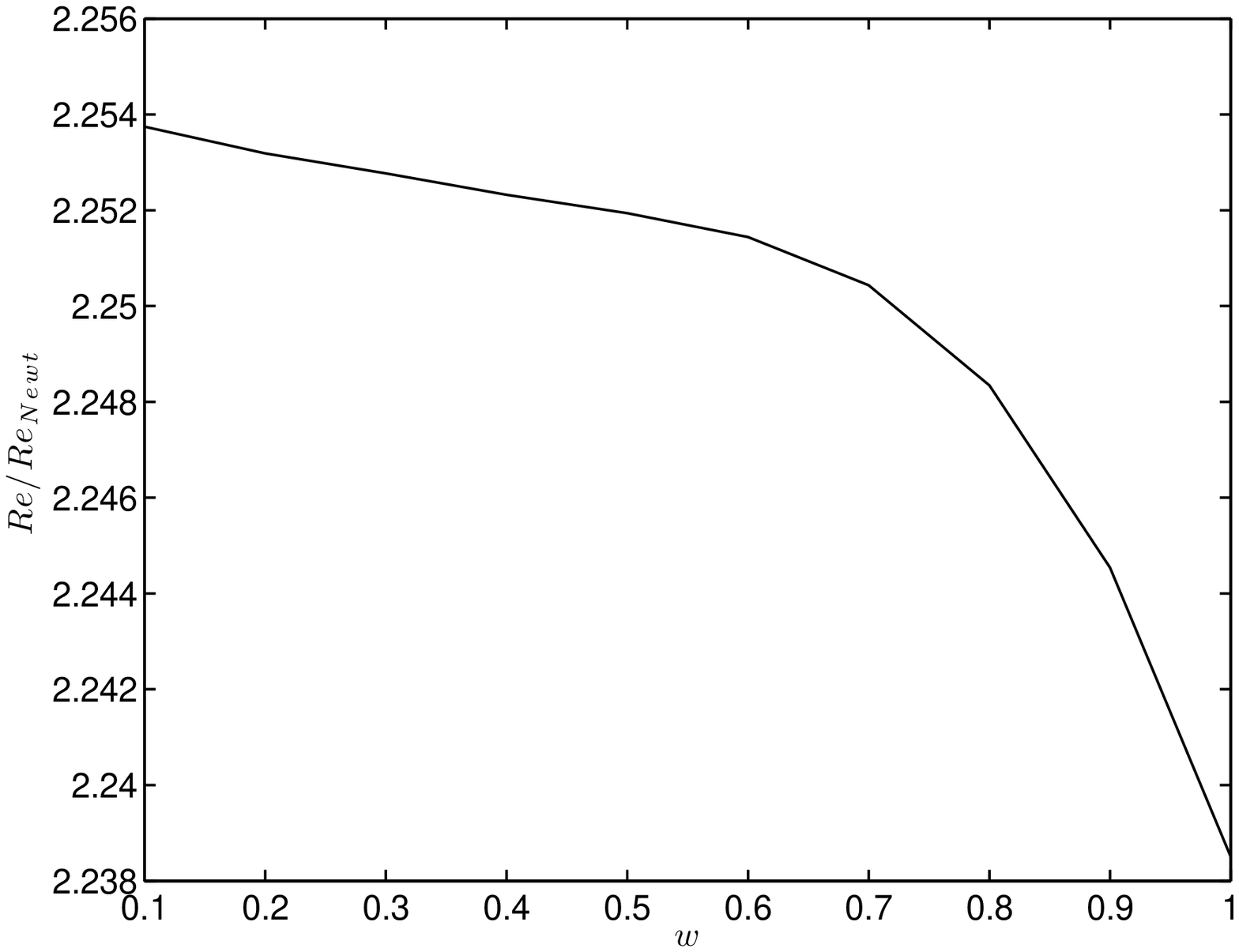}}
(b)\scalebox{0.5}{\includegraphics{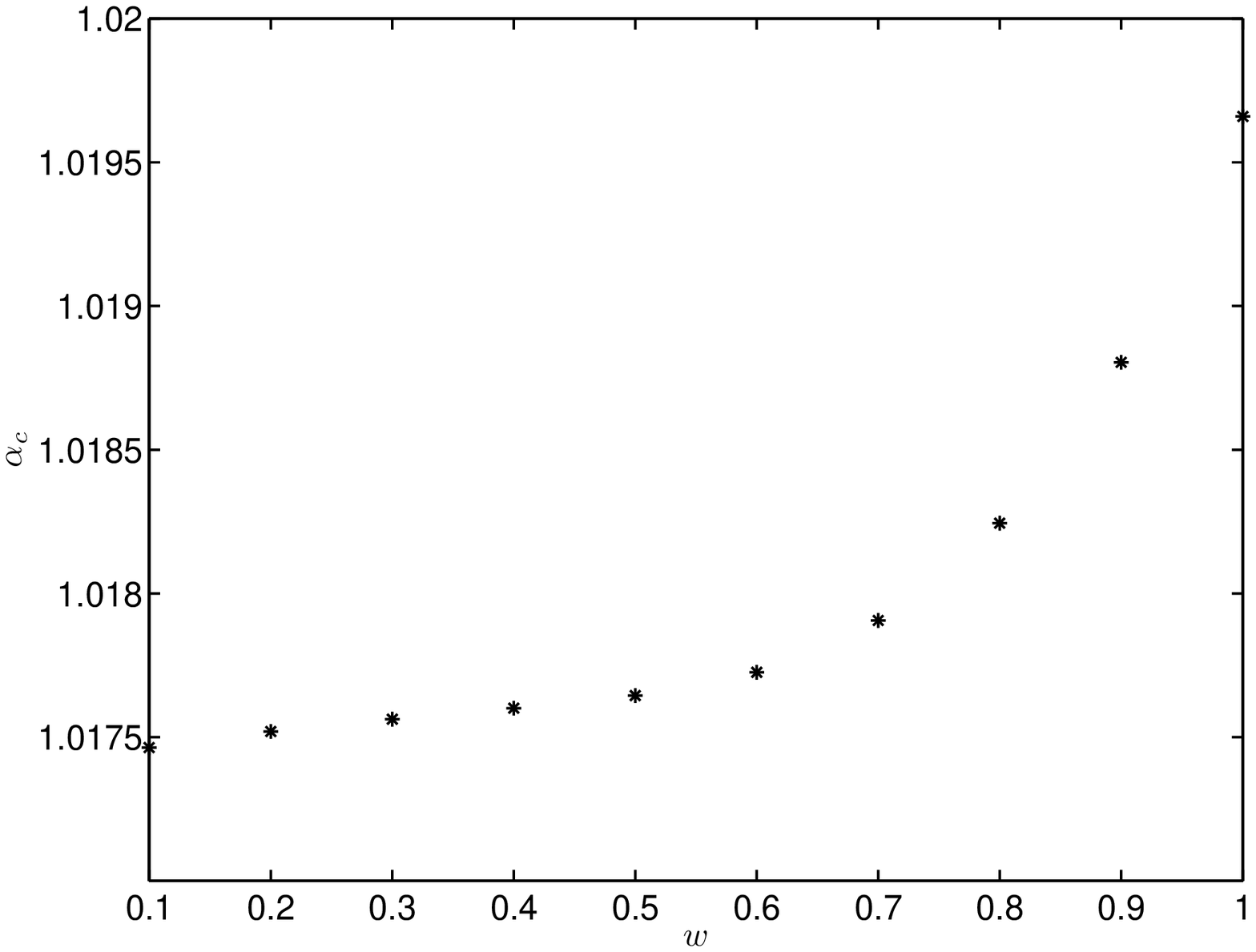}}
\end{center}
\caption{(a) Normalized  critical $Re$ with $Pl=1000$ and increasing $w$. (b) Critical wave number. Parameters for these calculations: $\epsilon=0.002$, $\elast=10^{-4}$, $m=2.5$, $K_r=K_d=0.3$ and $\mu_r=0.1$.}
\label{fig:resultspmmw}
\end{figure}

\begin{figure}
\begin{center}
(a)\scalebox{0.5}{\includegraphics{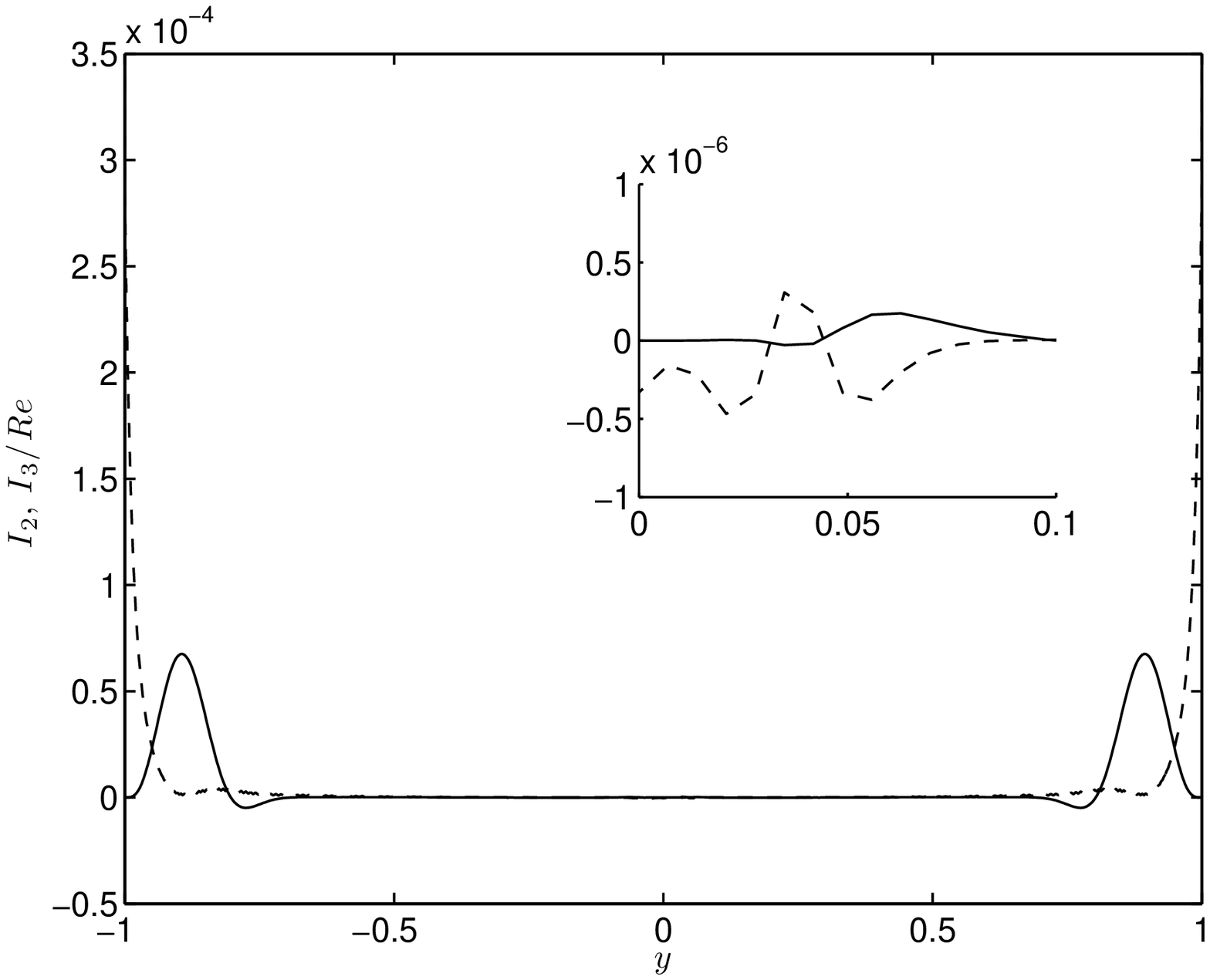}}
(b)\scalebox{0.5}{\includegraphics{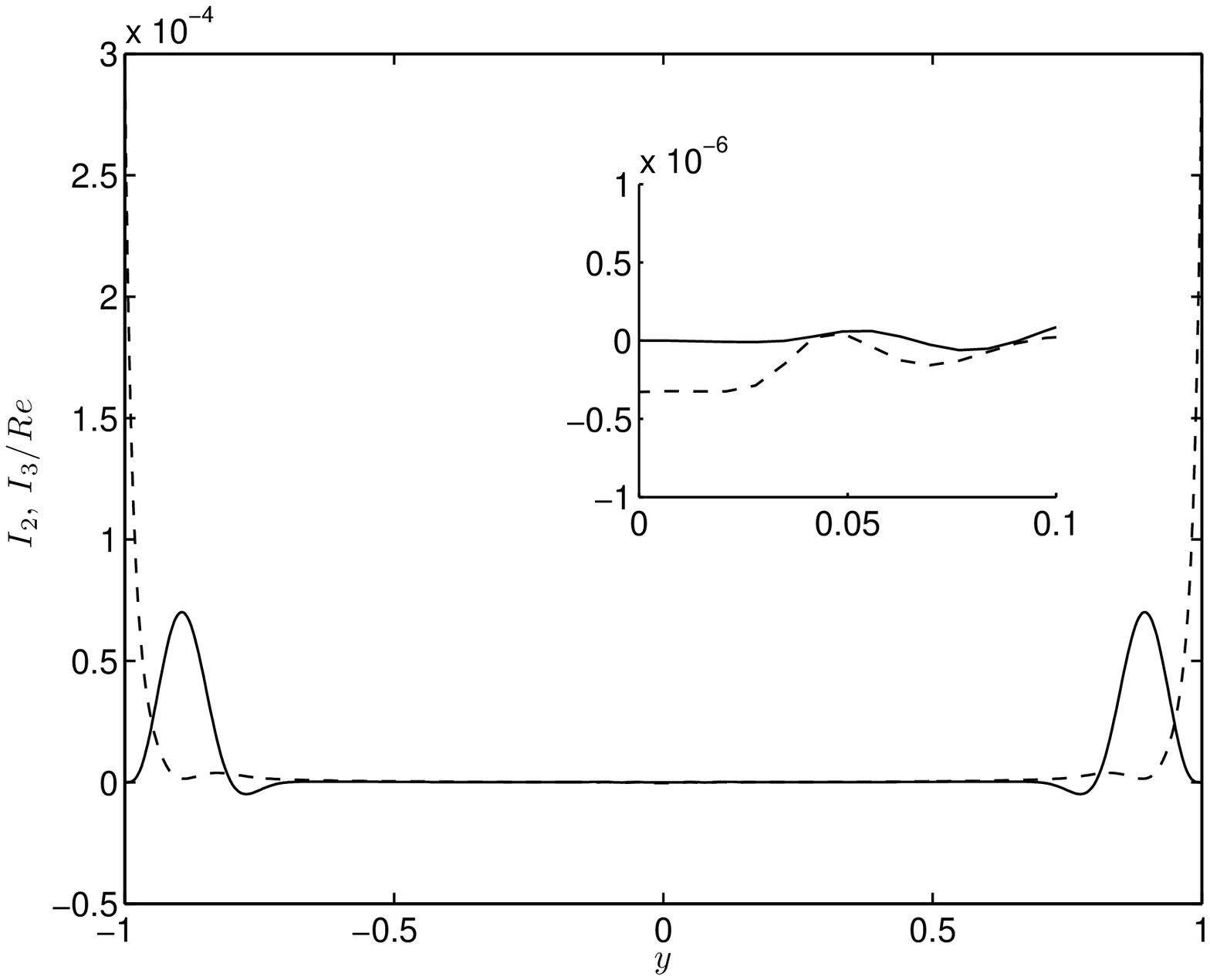}}
\end{center}
\caption{Energy budget for $Re=13364.41$ and $Pl=1000$. (a)Elastoviscoplastic model with $\elast=10^{-3}$, $\Gamma^+=7.356\times10^{-3},~\Gamma^-=7.03\times10^{-3}$ (b) Elastoviscoplastic model with $w=1$, $\Gamma^+=7.436\times10^{-3},~\Gamma^-=7.037\times10^{-3}$. Parameters for these calculations: $\epsilon=0.002$, $m=2.5$, $K_r=K_d=0.3$ and $\mu_r=0.1$.}
\label{fig:energybudget2}
\end{figure}

Now we explore the effects that increasing elasticity will have on the stability of the flow. In Figure \ref{fig:resultspmmeps}a we present the normalized critical Reynolds number for increasing $\elast$. Note that the change is minimal, the same happens for the critical wave number shown in Figure \ref{fig:resultspmmeps}b. This is not surprising either, elastic forces are confined to regions of very low shear-rates where the value of the nonlinear viscosity is very large, thus an increase on $\elast$ does not make any difference. This is also seen in Figure \ref{fig:energybudget2}a, where $\elast=10^{-3}$ and $\Gamma^+_{ev}$ and $\Gamma^-_{ev}$ are almost unchanged with respect to the case $\elast=10^{-4}$. In Figure \ref{fig:resultspmmw}a-b we show the effects of increasing the parameter $w$ which in turns makes the solid-fluid region wider. Note that for small values of $w$ things remain more less unchanged but as soon as $w$ gets closer to 1 we see a steep decrease on the critical Reynolds number. This again is not too surprising, by increasing $w$ the region where solid-fluid coexist stretches towards the wall. When $w=1$ we have that both $\Phi$ and $T_{xx}$ are different from zero at the wall. This is clearly seen in Figure \ref{fig:energybudget2}b where the normalized production of energy $\Gamma^+_{ev}$ has increased in the critical layer.

%%%%%%%%%%%%%%%%%%%%%%%%%%%%%%%%%%%%%%%%%%%%%%%%%%%%%%%%%%%%%%%%%%%%%%%%
%%%%%%%%%%%%%%%%%				Discussion									%%%%%%%%%%%%%%%%%%%
%%%%%%%%%%%%%%%%%%%%%%%%%%%%%%%%%%%%%%%%%%%%%%%%%%%%%%%%%%%%%%%%%%%%%%%%
\section{Discussion}

We have studied linear stability analysis of plane Poiseuille flow of an elastoviscoplastic fluid. Our results show that there is an increase in the critical Reynolds number for two-dimensional perturbations as the yield stress increases. The main difference between our study and previous works is that we consider a rather new yielding scenario recently suggested in  \cite{Burghelea09} according to which the transition from solid to fluid regime is not direct but mediated by a solid-fluid phase coexistence regime within which the material behaves as a viscoelastic fluid. Thus, we were interested in understanding how the presence of elasticity modifies the hydrodynamic stability of the flow. 
The relevance of a particular yielding scenario to the stability problem might not be obvious at a first glance as the yielding typically occurs at low Reynolds numbers (the experiments described in  \cite{Burghelea09} were performed at Re<1) whereas the loss of hydrodynamic stability emerges at significantly higher Re. However, it has been demonstrated experimentally in \cite{bulent_crap} that the transition to turbulence in the pipe flow of a viscoplastic fluid (Carbopol 940) is simultaneous with the breakdown of the unyielded plug initially located around the centerline of the pipe. These results suggested us that the loss of the hydrodynamic stability and the yielding are in fact connected, which motivated us to revisit the stability problem in the context of the yielding scenario suggested in  \cite{Burghelea09}. These results are in a close agreement with the ones of a regularized viscoplastic model, this is not surprising due to the fact that the viscoelastic core is confined to a region away from the wall. We have also shown that the existence of a region where solid and fluid structure coexist destabilizes the flow.

We have chosen the regularized Casson model for the viscosity function but we suspect that the stability behaviour is qualitatively similar for other models such as the regularized Hershel-Bulkley and Bingham models, and therefore that our results are quite generic. 

There are practical limitations to be acknowledged. First, the domain is a geometric idealisation of a planar geometry of large aspect ratio. Depending on the actual geometry it becomes necessary to consider other effects, e.g.~entry/start-up effects of the flow, curvature or imperfections of the walls, etc. On the other hand in the experimental setting, the point at which instability is actually observed is very sensitive to control of apparatus imperfections and the level of flow perturbations. For example, in Hagen-Poiseuille flow of Newtonian fluids one typically observes transition to turbulence starting for $Re \gtrsim 2000$. However, an experimental flow loop in Manchester UK produces stable laminar flows for $Re \approx 24,000$, \cite{Hof04,Peixinho06}, and stable flows have even been reported up to $Re \approx 100,000$, \cite{Pfenniger61}. This suggests that enhanced stability may be achieved experimentally, where predicted by the linear theory.

%%%%%%%%%%%%%%%%%%%%%%%%%%%%%%%%%%%%%%%%%%%%%%%%%%%%%%%%%%%%%%%%%%%%%%%%
%%%%%%%%%%%%%%%%%				Appendix  									%%%%%%%%%%%%%%%%%%%
%%%%%%%%%%%%%%%%%%%%%%%%%%%%%%%%%%%%%%%%%%%%%%%%%%%%%%%%%%%%%%%%%%%%%%%%
\appendix

\section{Bounds for Orr-Somerfeld and Squire modes}

%-----------------------------------------------------
\subsection{Proof of Theorem \ref{thm:nn-squire}}\label{App:Squire-proof}

\begin{proof}
The homogeneous modified Squire equation is obtained by setting $v\equiv0$ in \eqref{eq:modified-os-sq}. This gives
\begin{equation}\label{pf:nn-squire1}\begin{aligned}
c\eta=&U\eta+\frac{i}{\alpha{Re}}\mu(D^2-k^{2})\eta + \frac{i}{\alpha{Re}}D\mu D\eta\\
&+\frac{i}{\alpha{Re}}\frac{\beta^{2}}{k^{2}}D[(\mu_{t}-\mu)D\eta].
\end{aligned}\end{equation}
Multiplying \eqref{pf:nn-squire1} by the conjugate $\eta^*$, integrating by parts over $y\in[-1,1]$ and using boundary conditions we get:
\begin{equation}\label{pf:nn-squire3}
c\int|\eta|^{2}\ dy=\int{}U|\eta|^{2}\ dy+\frac{i}{\alpha{Re}}\int\left[\mu(|D \eta|^{2}-k^{2}|\eta|^{2})+\frac{\beta^{2}}{k^{2}}(\mu_{t}-\mu)|D\eta|^{2}\right]\ dy.
\end{equation}
Using the second mean value theorem for integrals and taking real and imaginary parts, we get:
\begin{equation}\label{pf:nn-squire-real}
c_{r}I_{0}^{2} = U(y_{1})I_{0}^{2},
\end{equation}
\begin{equation}\label{pf:nn-squire-imag}
c_{i}I_{0}^{2} = \frac{-1}{\alpha{Re}}\left[\left(1-\frac{\beta^{2}}{k^{2}}\right)\mu(y_{2})I_{1}^{2}+\mu(y_{3})k^{2}I_{0}^{2}+\mu_{t}(y_{4})\frac{\beta^{2}}{k^{2}}I_{1}^{2}\right],
\end{equation}
with $y_{i}\in[-1,1]$ to be the mean values. The first bound immediately comes from the fact that $U_{min}<U(y_{i})<U_{max}$. Note that for shear thinning fluids, $\mu(y_{i})>\min\mu>\min\mu_{t}>0$. We also have $\mu_{t}(y_{i})>\min\mu_{t}$. Thus, for $p=\min\mu_{t}>0$,
\begin{equation}\label{pf:squire4}\begin{aligned}
c_{i}I_{0}^{2} &< -\frac{1}{\alpha{Re}}\left[p\left(1-\frac{\beta^{2}}{k^{2}}\right)I_{1}^{2}+pk^{2}I_{0}^{2}+p\frac{\beta^{2}}{k^{2}}I_{1}^{2}\right]\\
&=-\frac{p}{\alpha{Re}}\left[I_{1}^{2} + k^{2}I_{0}^{2}\right]<0,
\end{aligned}\end{equation}
so all the Squire modes are damped since $I_{0}^{2}$ is positive. Dividing through by $I_{0}^{2}$ and applying our isoperimetric inequality $I_{1}^{2}/I_{0}^{2}\geq\pi^{2}/4$ given in Lemma \ref{thm:poincare} we get our result as required. \end{proof}

%--------------------------
\subsection{Proof of Theorem \ref{thm:nn-joseph}}\label{App:Orr-Somm-proof}
\begin{proof}

Using the fact that the Squire modes are always damped, we multiply \eqref{Orr-Somreg} by the conjugate $v^*$, integrate by parts, use boundary conditions, take real and imaginary parts and use the second mean value theorem for integrals to get:
\begin{equation}\label{pf:nn-joseph-imag1}
c_{i}(I_{1}^{2}+\alpha^{2}I_{0}^{2})=(Q-Q^*)-(\alpha{Re})^{-1}4\alpha^{2}\hat{\mu}I_{1}^{2} -(\alpha{Re})^{-1}\hat{\mu}_{t}\int|v^{\prime\prime}+\alpha^{2}v|^{2}\ dy,
\end{equation}
\begin{equation}\label{pf:nn-joseph-real}
c_{r}(I_{1}^{2}+\alpha^{2}I_{0}^{2})=\int\left[{U}|v^{\prime}|^{2}+(\alpha^{2}U+U^{\prime\prime}/2)|v|^{2}\right]\ dy,
\end{equation}
where $Q=(i/2)\int{U^{\prime}}v^{*'}v\ dy$, $\hat{\mu}=\mu(y_1)$, $\hat{\mu}_t=\mu_t(y_2)$ with $y_1,~y_2\in[-1,1]$. We can expand the last integral of the right hand side of \eqref{pf:nn-joseph-imag1} and integrate by parts. This gives
\begin{equation}\label{pf:nn-joseph-imag}
c_{i}(I_{1}^{2}+\alpha^{2}I_{0}^{2})=(Q-Q^*)-(\alpha{Re})^{-1}[4\alpha^{2}\hat{\mu}I_{1}^{2}-\hat{\mu}_{t}(I_{2}^{2}-2\alpha^{2}I_{1}^{2}-\alpha^{4}I_{0}^{2})].
\end{equation}
We will now proceed with the proof.

As before, note that $\hat\mu>\min\mu>\min\mu_{t}$, thus we divide by $(I_{1}^{2}+\alpha^{2}I_{0})$, using Cauchy-Schwartz inequality on the first term of the right hand side of \eqref{pf:nn-joseph-imag} and our definitions for $p$ and $q$, we have
\begin{equation}\label{pf:nn-joseph4}\begin{aligned}
c_{i}&<\frac{qI_{1}I_{0}}{I_{1}^{2}+\alpha^{2}I_{0}^{2}} -\frac{(\alpha{Re})^{-1}(4\alpha^{2}pI_{1}^{2}+p(I_{2}^{2}-2\alpha^{2}I_{1}^{2}-\alpha^{4}I_{0}^{2}))}{I_{1}^{2}+\alpha^{2}I_{0}^{2}}\\
&=\frac{qI_{1}I_{0}}{I_{1}^{2}+\alpha^{2}I_{0}^{2}} -\frac{(\alpha{Re})^{-1}p(I_{2}^{2}+2\alpha^{2}I_{1}^{2}+\alpha^{4}I_{0}^{2}))}{I_{1}^{2}+\alpha^{2}I_{0}^{2}}.
\end{aligned}\end{equation}
Clearly, equations \eqref{pf:nn-joseph-real} and \eqref{pf:nn-joseph4} are the same as equations (3) and (4a,b) in  \cite{Joseph68} respectively, with the inclusion of the positive constant $p$ in  \eqref{pf:nn-joseph4}. Therefore the proof follows in the same way as in  \cite{Joseph68} and our results hold. \end{proof}

\bibliography{pmmstab}
\bibliographystyle{amsplain}

\end{document}